\documentclass[prb,twocolumn,amsmath,amssymb,superscriptaddress,floatfix,nofootinbib]{revtex4-2}

\usepackage{graphicx}
\usepackage{dcolumn}
\usepackage{bm}
\usepackage{textcomp}
\usepackage{color}
\usepackage{amsmath}
\usepackage{amssymb}
\usepackage{xcolor}
\usepackage[colorlinks = true,
            linkcolor = blue,
            urlcolor  = red,
            citecolor = blue,
            anchorcolor = blue]{hyperref}
\usepackage{easyReview}
\usepackage{mathdots}
\usepackage{float}
\usepackage{lineno}
\usepackage{todonotes}
\usepackage{url}
\usepackage{array}
\usepackage{tabularx}
\usepackage{todonotes}
\usepackage{xcolor,colortbl}

\usepackage{lipsum, babel}

\definecolor{LightBlue}{rgb}{0.8,0.8,0.8}

\usepackage[normalem]{ulem}

\begin{document}

\title{Interlayer Pairing in Bilayer Nickelates}

\date{\today}

\author{Thomas A. Maier} \affiliation{Computational Sciences and Engineering Division, Oak Ridge National Laboratory, Oak Ridge, Tennessee 37831, USA}
\author{Peter Doak} \affiliation{Computational Sciences and Engineering Division, Oak Ridge National Laboratory, Oak Ridge, Tennessee 37831, USA}
\author{Ling-Fang Lin} \affiliation{Department of Physics and Astronomy, University of Tennessee, Knoxville, Tennessee 37996, USA}
\author{Yang Zhang} \affiliation{Department of Physics and Astronomy, University of Tennessee, Knoxville, Tennessee 37996, USA}
\author{Adriana Moreo} \affiliation{Department of Physics and Astronomy, University of Tennessee, Knoxville,  Tennessee 37996, USA} \affiliation{Materials Science and Technology Division, Oak Ridge National Laboratory, Oak Ridge, Tennessee 37831, USA}
\author{Elbio Dagotto} \affiliation{Department of Physics and Astronomy, University of Tennessee, Knoxville,  Tennessee 37996, USA} \affiliation{Materials Science and Technology Division, Oak Ridge National Laboratory, Oak Ridge, Tennessee 37831, USA}

\begin{abstract}
The discovery of $T_c\sim 80$~K superconductivity in pressurized La$_3$Ni$_2$O$_7$ has launched a new platform to study high-temperature superconductivity. Using non-perturbative dynamic cluster approximation quantum Monte Carlo calculations, we characterize the magnetic and superconducting pairing behavior of a realistic bilayer two-orbital Hubbard-Hund model of this system that describes the relevant Ni $e_g$ states with physically relevant interaction strengths. We find a leading $s^\pm$ superconducting instability in this model and show that this state primarily arises from interlayer pairing in the $d_{3z^2-r^2}$ orbital that is driven by strong interlayer spin-fluctuations in that orbital. These results provide non-perturbative evidence supporting the picture that a simple single-orbital bilayer Hubbard model for the Ni $d_{3z^2-r^2}$ orbital provides an excellent low-energy effective description of the superconducting behavior of La$_3$Ni$_2$O$_7$.

\end{abstract}

\maketitle

\section*{Introduction}

Superconductivity in pressurized La$_3$Ni$_2$O$_7$ \cite{sun_signatures_2023, hou_emergence_2023} has been widely addressed in bilayer two-orbital Hubbard models that account for the $e_g$ manifold ($d_{x^2-y^2}$ and $d_{3z^2-r^2}$ orbitals) of the Ni-$d$ states near the Fermi level in these systems \cite{PhysRevLett.131.126001,PhysRevLett.131.236002,PhysRevB.108.L201121,zhang_structural_2024,PhysRevLett.132.106002,gu_effective_2025,xi_transition_2025,YangYf_2023_Interlayer}. Due to the complexity of the electronic structure, most of these studies have used perturbative, either weak-coupling \cite{PhysRevB.108.L201121,PhysRevLett.131.236002,zhang_structural_2024,BotzelS_2024_Theory,xia_sensitive_2025,PhysRevLett.132.106002,xi_transition_2025,Heier_G_2024_Competing,YangQG_2023_Possible,gu_effective_2025} or strong-coupling \cite{Lu_C_2024_Interlayer,YangYf_2023_Interlayer,liaoZ_2023_Electron,Jiang_2024} approaches. Depending on details in the model parameters \cite{xia_sensitive_2025} and the type of the approximation, these studies have found $s^\pm$, $d_{x^2-y^2}$-, and $d_{xy}$-wave superconducting states.

Optical studies, however, show evidence that La$_3$Ni$_2$O$_7$ is characterized by moderately strong electronic correlations \cite{liu_electronic_2024,geisler_optical_2024} in the intermediate coupling regime. Consistent with this, recent density functional theory and constrained random-phase approximation (RPA) calculations indeed find that the Hubbard $U$ interaction on the Ni $e_g$ orbitals is approximately of the same size as the electronic bandwidth \cite{yue2025correlatedelectronicstructuresunconventional}. This raises the question of whether perturbative weak- or strong-coupling approaches can accurately characterize the nature of pairing in these systems, and underscores the need for non-perturbative methods to address this question. While numerical methods such as density matrix renormalization group and tensor networks \cite{Shen_2023,QuXZ_2024_Bilayer,schlomer_superconductivity_2024,PhysRevB.109.045154,PhysRevB.109.L201124}, auxiliary-field Monte Carlo \cite{PhysRevB.108.L140504}, and cluster dynamical mean-field theory (CDMFT) \cite{zheng_s_ifmmodepmelsetextpmfi-wave_2025,tian_correlation_2024} have been used to provide a non-pertubative picture, they are typically based on reduced effective t-J models, simplified interactions, or quasi-one dimensional lattice geometries.

Here, by using state-of-the-art dynamical cluster approximation (DCA) quantum Monte Carlo \cite{maier_quantum_2005} calculations for a realistic bilayer two-orbital model on a two-dimensional lattice \cite{zhang_structural_2024} with physically relevant interaction parameters, we examine what this model tells us about the pairing mechanism in the bilayer La$_3$Ni$_2$O$_7$ compound. We find a leading $s^\pm$ superconducting instability in this model at a temperature $T\sim 100$~K close to the experimentally observed $T_c\sim 80$~K. We show that this instability arises from interlayer electron pairing on neighboring sites in the top and bottom layers, primarily in the $d_{3z^2-r^2}$ orbital. By analyzing the spin correlations in the model and correlating their behavior with the $s^\pm$ pairing interaction strength, we demonstrate that the pairing is primarily driven by interlayer spin fluctuations arising from the $d_{3z^2-r^2}$ orbital. Hence, these results support a picture of a single-orbital bilayer Hubbard model \cite{maier_pair_2011} for the $d_{3z^2-r^2}$ orbital providing an appropriate low-energy effective description of the pairing behavior of La$_3$Ni$_2$O$_7$. While this relation was discussed before by Sakakibara \textit{et al.} \cite{PhysRevLett.132.106002} in the context of a perturbative weak-coupling analysis, our state-of-the-art numerical results prove the validity of this picture by treating the strong interactions in this model non-perturbatively.

\section*{Results}
\begin{figure}[ht]
  \includegraphics[width=0.48\textwidth]{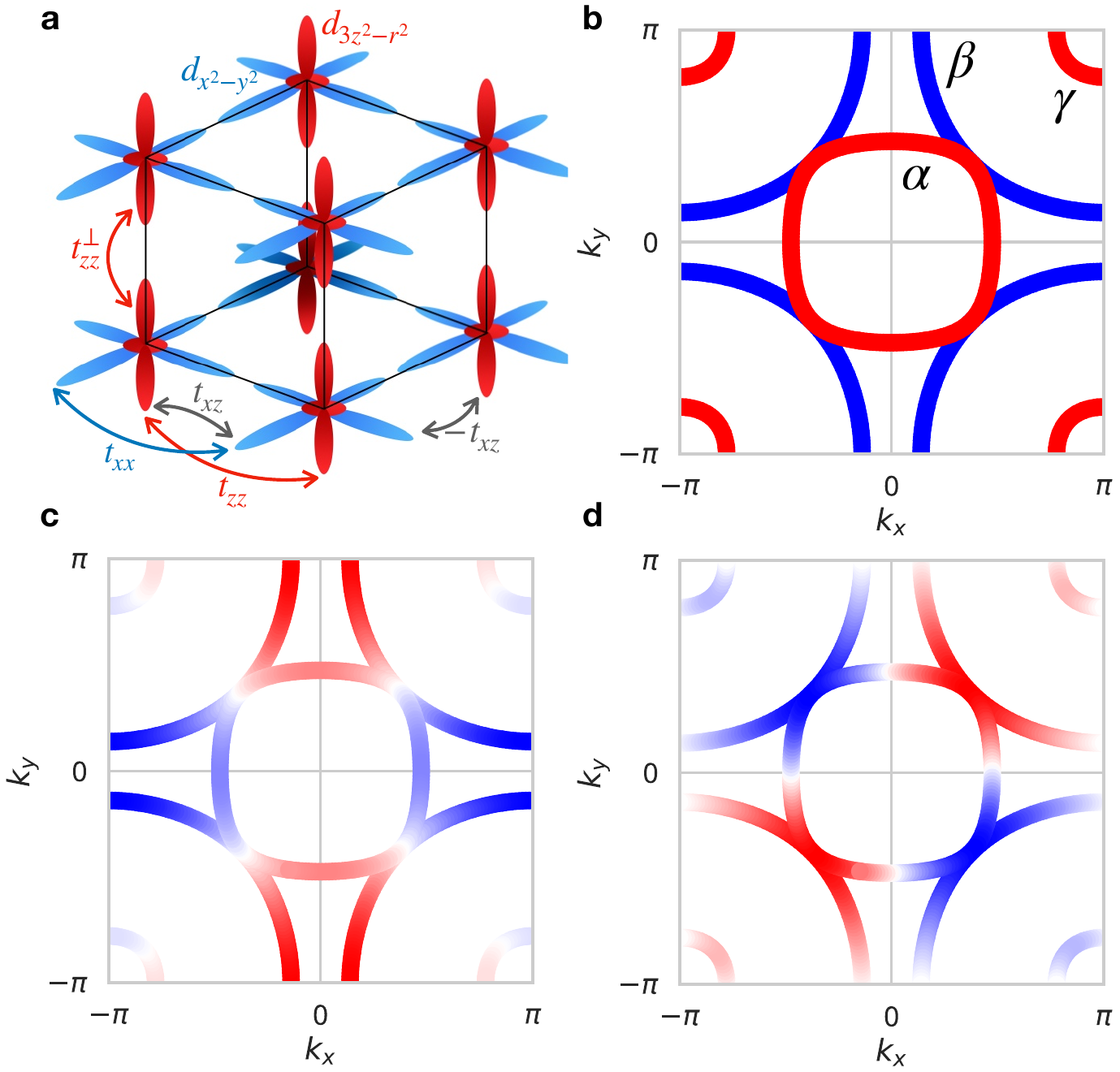}
  \caption{Illustration of the bilayer two-orbital $d_{x^2-y^2}$ (blue), $d_{3z^2-r^2}$ (red) model (a) with nearest neighbor hopping parameters $t_{xx}=-0.515$, $t_{zz}=-0.110$, $t_{xz}=0.243$, and $t^\perp_{zz}=0.666$ taken from Ref.~\cite{zhang_structural_2024} for 25~GPa pressure.  Its non-interacting Fermi surface consistst of $\alpha$, $\beta$, and $\gamma$ sheets. The superconducting order parameter form factors listed in Tab.~\ref{tab:gk} for $s^\pm$ (b), $d_{x^2-y^2}$ (c), and $d_{xy}$ (d) states are illustrated by red (positive) and blue (negative) colors. \label{fig:1}}
\end{figure}
The model we consider was introduced in Ref.~\cite{zhang_structural_2024} and is illustrated in Fig.~\ref{fig:1}a. Its Hamiltonian is given in the Methods section. It includes the two Ni-$3d$ orbitals, $d_{x^2-y^2}$ and $d_{3z^2-r^2}$, that have been found to account for the low-energy electronic structure of the bilayer nickelates. These two orbitals are located on a two-dimensional (2D) square bilayer lattice, and the model includes nearest neighbor hopping, both intra- and inter-layer. It also includes  intra-orbital Coulomb $U$, inter-orbital Coulomb $U'$, and Hund's rule coupling $J$ interactions, which we set to $U=3$~eV, $U'=2$~eV, and $J=0.5$~eV (see also Methods section).

\subsection*{Pair-field susceptibility}
We start by discussing results for the pair-field susceptibility defined as
\begin{equation}\label{eq:Palpha}
P_\alpha(T) = \int_0^\beta d\tau \langle {\cal T}_\tau \Delta^{\phantom\dagger}_\alpha(\tau)\Delta^{\dagger}_\alpha(0)\rangle
\end{equation}
with the pair operator
\begin{equation}
\Delta^{\dagger}_\alpha = \frac{1}{\sqrt{N}}\sum_{\bm k,\ell\ell'} g^{\ell\ell'}_\alpha(\bm k) c^\dagger_{\bm k \ell\uparrow}c^\dagger_{-\bm k \ell'\downarrow}\,.
\end{equation}
Here we have used the momentum space Fourier representation with wave-vector ${\bm k}=(k_x, k_y, k_z)$ with $k_z=0$ or $\pi$ representing bonding and anti-bonding combinations, respectively, of the two layers, $c^\dagger_{\bm k\ell\sigma}=1/\sqrt{N}\sum_{i} c^\dagger_{i\ell\sigma}e^{i{\bm k}{\bm r_i}}$, and $g_\alpha^{\ell\ell'}(\bm k)$ is a symmetry form-factor. The singlet form-factors we will use are listed in Tab.~\ref{tab:gk} and illustrated in Fig.~\ref{fig:1}b, c, and d. They correspond to local inter-layer $s^\pm$ pairs and in-plane $d_{x^2-y^2}$ and $d_{xy}$ pairs. As seen in Fig.~\ref{fig:1}b, the $s^\pm$ state is positive on the $\beta$ sheet and negative on the $\alpha$ and $\gamma$ pockets. This sign change arises from the difference in bonding/antibonding character of the Fermi surface states on the $\beta$ and $\gamma$ pockets \cite{zhang_structural_2024}. Since the $d_{x^2-y^2}$ and $d_{xy}$ states are in-plane states, they are in-phase across all the Fermi surface sheets. In all cases, we only consider intra-orbital pairing. We note that the $N_c=2$ site cluster calculation can already address the $s^\pm$ state, since it contains the two sites (one in the top and one in the bottom layer) that are involved in this pair. It may be regarded as the mean-field solution for this state. Additionaly, the $N_c=8$ site cluster can also address the $d_{x^2-y^2}$ pairing state, but not the $d_{xy}$ state. The $N_c=16$ cluster can resolve all the symmetries we consider.

\begin{table}[h]
{\renewcommand{\arraystretch}{1.2}
\begin{center}
\begin{tabular}{ll}
\hline
$\alpha$ & $g^{\ell\ell'}_\alpha(\bm k)$\\[0.5ex]
\hline
$s^\pm$ & $\cos k_z \delta_{\ell\ell'}$\\
$d_{x^2-y^2}$ & $(\cos k_x - \cos k_y) \delta_{\ell\ell'}$\\
$d_{xy}$ & $\sin k_x \sin k_y \delta_{\ell\ell'}$\\
\end{tabular}
\end{center}
}
\caption{Singlet order parameter form-factors used for the calculation of the pair-field susceptibility in Eq.~(\ref{eq:Palpha}). \label{tab:gk}}
\end{table}

\begin{figure}[h]
  \includegraphics[width=0.5\textwidth]{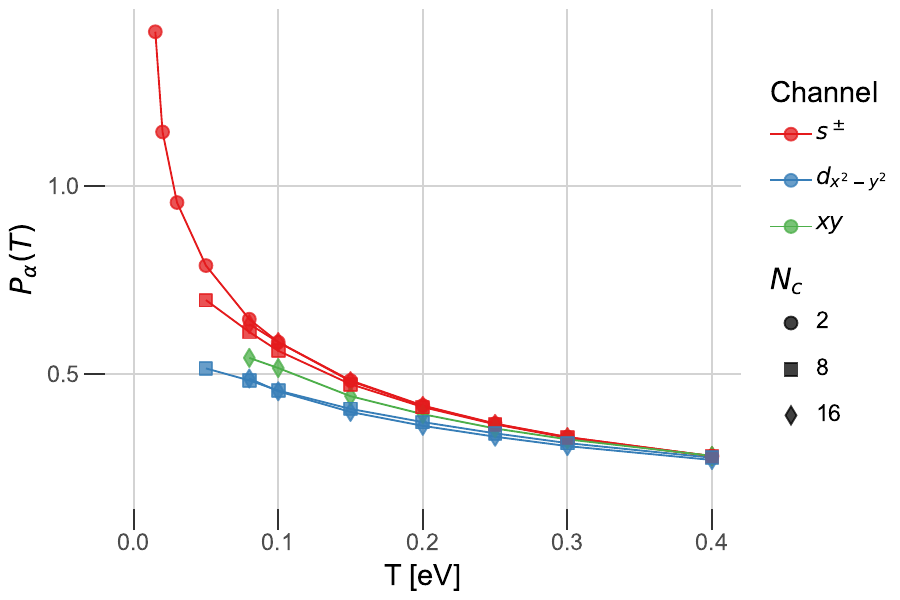}
  \caption{The pair-field susceptibility versus temperature for the two-orbital bilayer 327-LNO model with $U=3$~eV, $U'=2$~eV, and $J=0.5$~eV for different symmetry channels $\alpha$ and DCA cluster sizes $N_c$. The leading $s^\pm$ symmetry channel has a mean-field superconducting transition near $T\sim 0.01$~eV $\sim 100$~K. \label{fig:2}}
\end{figure}

Fig.~\ref{fig:2} shows the temperature dependence of the pair-field susceptibility $P_\alpha(T)$ for the the different symmetry channels and cluster sizes. Earlier RPA calculations for the same model found a leading $s^\pm$ superconducting state, with a sub-leading $d_{x^2-y^2}$ state \cite{zhang_structural_2024}. Consistent with this, one sees in Fig.~\ref{fig:2} that the $s^\pm$ susceptibility is rising the most rapidly at low temperatures, followed by the $d_{xy}$ and $d_{x^2-y^2}$ susceptibilities. The leading $s^\pm$ channel shows divergent behavior for the $N_c=2$ site cluster indicating an instability near $T\sim 0.01$~eV $\sim 100$~K. This result is consistent with the CDMFT study in Ref.~\cite{zheng_s_ifmmodepmelsetextpmfi-wave_2025} for a similar but different model and simplified interaction term with only a $U$ term on the $d_{3z^2-r^2}$ orbital. The $N_c=8$ results fall below the $N_c=2$ results, while the $N_c=16$ results are very close to the $N_c=2$ results for the temperatures we can access. Fluctuations in the cluster size of this sort are expected, since certain clusters, such as $N_c=8$, may overestimate phase fluctuations and therefore underestimate the pair-field susceptibility \cite{maier_systematic_2005}. Lower temperatures are inaccessible for the larger cluster sizes due to the Fermion sign problem. Nevertheless, based on these results, we conclude that the normal state of this model has an instability to an $s^\pm$ superconducting state at a temperature near $T\sim 0.01$~eV $\sim 100$~K.

\begin{figure}[ht!]
  \includegraphics[width=0.48\textwidth]{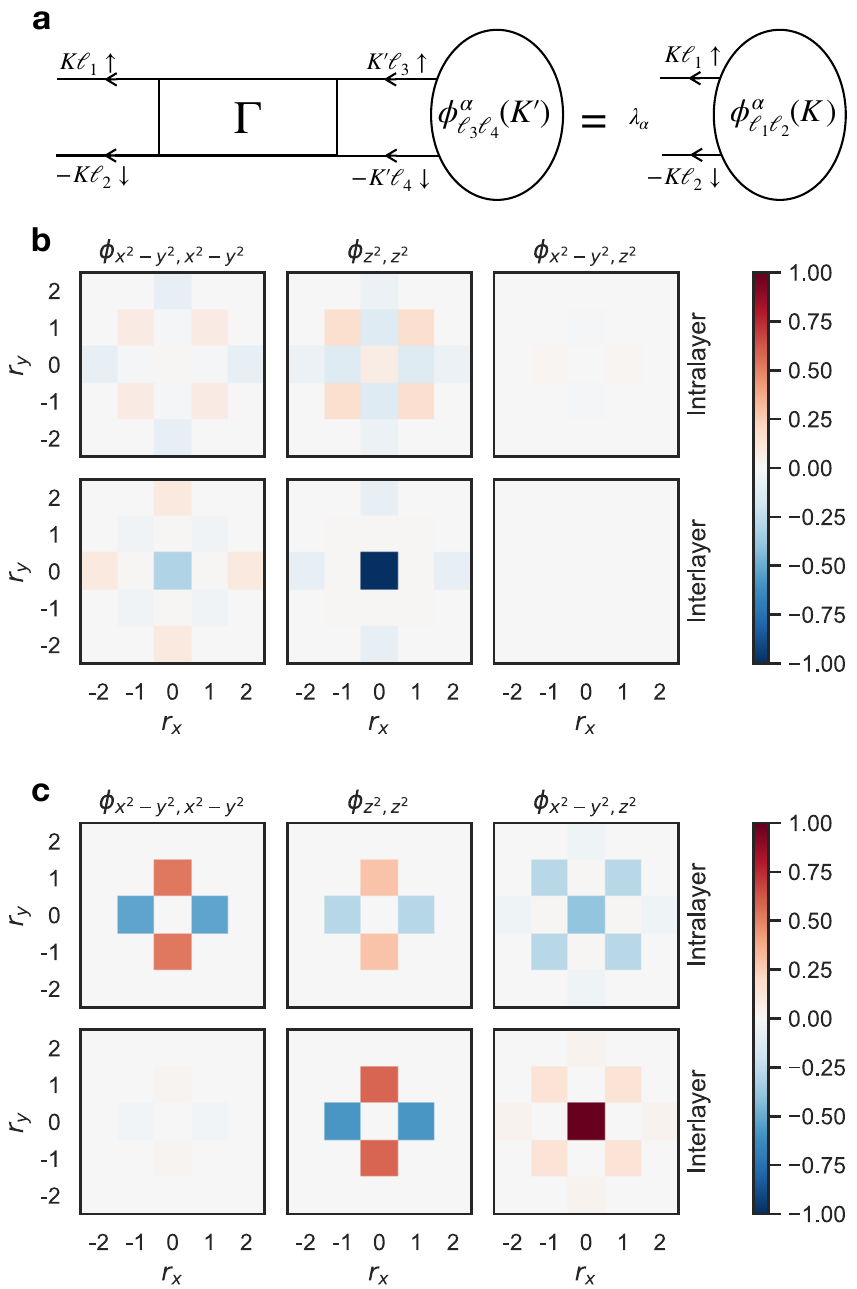}
  \caption{The eigenvalues $\lambda_\alpha$ and eigenvectors $\phi^\alpha_{\ell_1\ell_2}(K)$ of the interaction enhanced part of the pair-field susceptibility shown in panel (a) contain information about the leading pair structures. Panels (b) and (c) plot the spatial, orbital, and layer structure of the Fourier-transform of $\phi^\alpha_{\ell_1\ell_2}(K)$ to real space, $a^\alpha_{\ell_1\ell_2}(\bm r)$, for the leading $s^\pm$ and sub-leading $d_{x^2-y^2}$ pair correlations, respectively. Results are for the $N_c=16$ site cluster for $U=3$~eV, $U'=2$~eV, $J=0.5$~eV and temperature $T=0.08$~eV. \label{fig:3}}
\end{figure}

\subsection*{Orbital and real-space pair structure}
As illustrated in Fig.~\ref{fig:3}a and detailed in the Supplemental Material, the divergent interaction driven part of the pair-field susceptibility has the form $GG\,\Gamma\,GG$ with $G$ the single-particle Green's function and $\Gamma$ the reducible vertex. The leading eigenvalues $\lambda_\alpha$ and eigenvectors $\phi^\alpha_{\ell_1\ell_2}(K)$ of this quantity give information about the frequency and momentum/real-space structure of the pair correlations that give rise to the enhancement of the pair-field susceptibility. Since we are only interested in spin-singlet, even frequency correlations, we integrate this quantity over frequency as detailed in the Supplemental Material, and show in Fig.~\ref{fig:3}b and c the real-space, layer, and orbital structure of the two leading eigenvectors, $\phi^\alpha_{\ell_1\ell_2}(\bm r)=1/N_c\sum_{\bm K}\phi^\alpha_{\ell_1\ell_2}(\bm K)e^{i{\bm K}{\bm r}}$ for $\alpha=d_{x^2-y^2}$ and $d_{xy}$, respectively. For the leading $s^\pm$ eigenvector plotted in panel (b) of Fig.~\ref{fig:3}, the largest contribution arises from the $d_{3z^2-r^2}$ orbital and has inter-layer and local in-plane character. This corresponds to the $\cos k_z$ form factor in momentum-space. In addition, there is also a much smaller, inter-layer $\cos k_z$ contribution from the $d_{x^2-y^2}$ orbital, but only negligible intra-layer nearest- and next-nearest neighbor contributions for both orbitals. Inter-orbital pair correlations do not play a role for this state. The subleading $d_{x^2-y^2}$ pairing state shown in panel (c) also has a significant inter-layer $d_{3z^2-r^2}$ contribution, in this case with $d_{x^2-y^2}$ spatial structure. In contrast to the $s^\pm$ state in panel b, the $d_{x^2-y^2}$ orbital contributes strong intra-layer but no inter-layer correlations. Another difference is the presence of strong inter-orbital pair correlations, which have opposite signs for intra- and inter-layer contributions. These contributions necessarily have $s$-wave spatial structure, which, together with the overall $d_{x^2-y^2}$ orbital structure of the inter-orbital pairs, gives the net $d_{x^2-y^2}$-wave pair structure.

Notably, we find that the first state with $d_{xy}$-wave structure (not shown) has a significantly smaller eigenvalue than the $s^\pm$ and $d_{x^2-y^2}$ states shown in Fig.~\ref{fig:3}. This means that the $d_{xy}$-wave pair-field correlations shown in Fig.~\ref{fig:2}, even though they are larger than the $d_{x^2-y^2}$-wave correlations, arise mostly from the leading (zeroth order in the interaction) contribution to $P_{d_{xy}}$ and are not significantly enhanced by the interactions and therefore not divergent at lower temperatures.

\begin{figure}[ht]
  \includegraphics[width=0.48\textwidth]{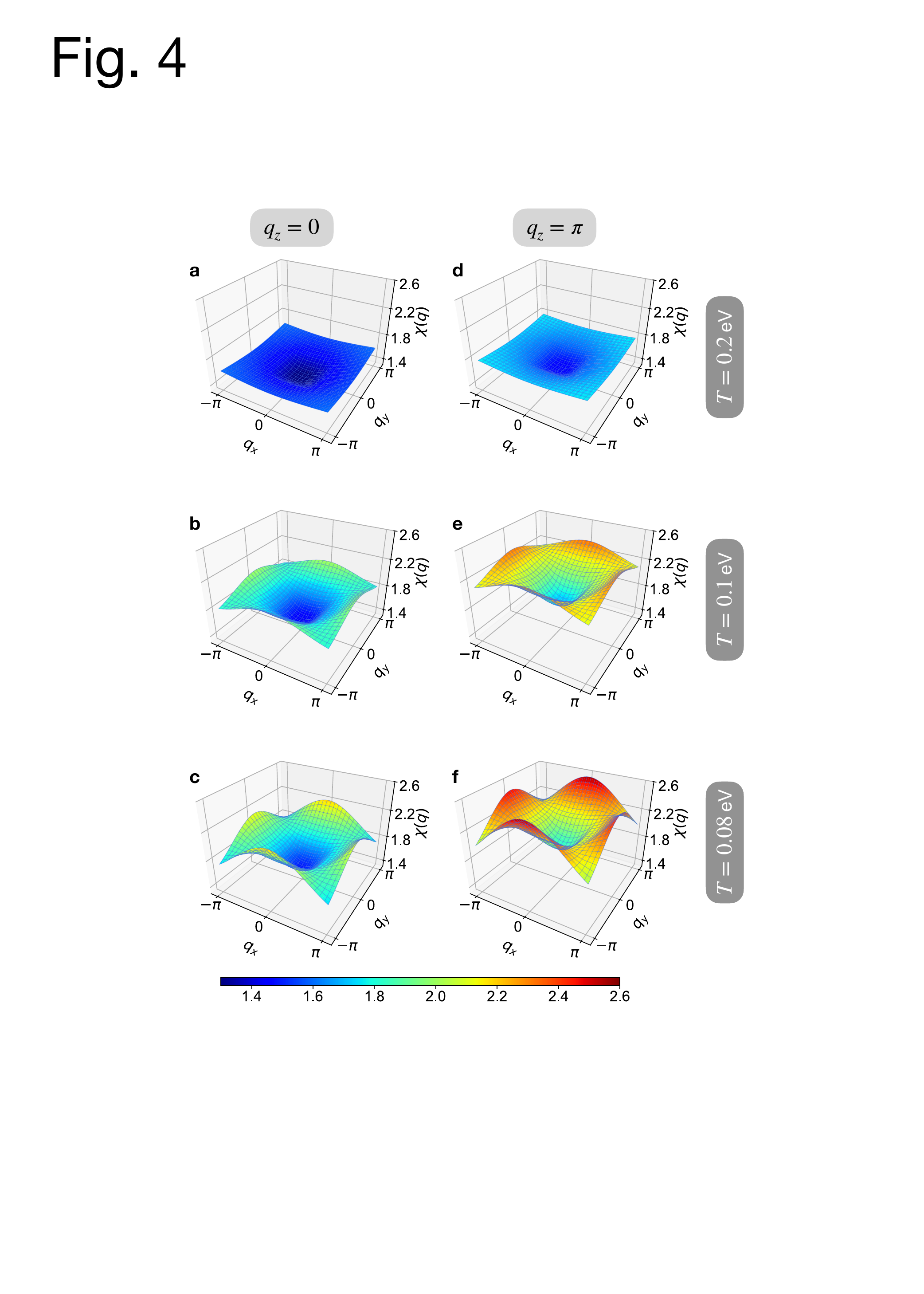}
  \caption{In-plane $(q_x, q_y)$ dependence of the cluster magnetic susceptibility $\chi(\bm q)$ for $q_z=0$ (left column) and $q_z=\pi$ (right column) for three different temperatures as indicated on the right. As the temperature is lowered, $\chi(\bm q)$ develops peaks for $\bm q=(\pi,0,\pi)$ and symmetry-related wavevectors. Results are for $U=3$~eV, $U'=2$~eV, and $J=0.5$~eV.  \label{fig:4}}
\end{figure}

\subsection*{Magnetic susceptibility}
Spin fluctuations have been argued to play an important role in mediating the pairing in unconventional superconductors like the nickelates. It is therefore interesting to study whether their structure is consistent with the observed pairing correlations. The zero frequency magnetic susceptibility is given by
\begin{equation} \label{eq:chis}
    \chi(\bm q) = \sum_{\ell_1\ell_2}\int_0^\beta d\tau \langle m^-_{\bm q}(\tau) m^\dagger_{\bm q}(0)\rangle
\end{equation}
with $m_{\bm q\ell}^\dagger = 1/\sqrt{N}\sum_{\bm k} d^\dagger_{\bm k+\bm q\ell\uparrow}d^{\phantom\dagger}_{\bm k\ell\downarrow}$ and $\bm q=(q_x,q_y,q_z)$ with $q_z=0$ and $\pi$ the even and odd layer combinations, respectively. We plot in Fig.~\ref{fig:4} the temperature dependence of $\chi(\bm q)$ calculated on the $N_c=16$ cluster and interpolated in $(q_x, q_y)$. For all temperatures, the inter-layer antiferromagnetic ($q_z=\pi$) correlations are stronger than the inter-layer ferromagnetic ($q_z=0$) correlations. As the temperature is lowered, peaks develop for in-plane $\bm q=(\pi,0)$ and symmetry-related points for both $q_z=0$ and $q_z=\pi$, so that at the lowest temperature, the strongest response is obtained for $\bm q=(\pi,0,\pi)$. This corresponds to in-plane striped (antiferromagnetic along $q_x$, ferromagnetic along $q_y$) and inter-layer antiferromagnetic correlations. These results are consistent with earlier weak coupling RPA calculations for this model in Ref.~\cite{zhang_structural_2024,PhysRevB.108.L180510}.

\subsection*{Pairing interaction}
\begin{figure}[h]
  \includegraphics[width=0.48\textwidth]{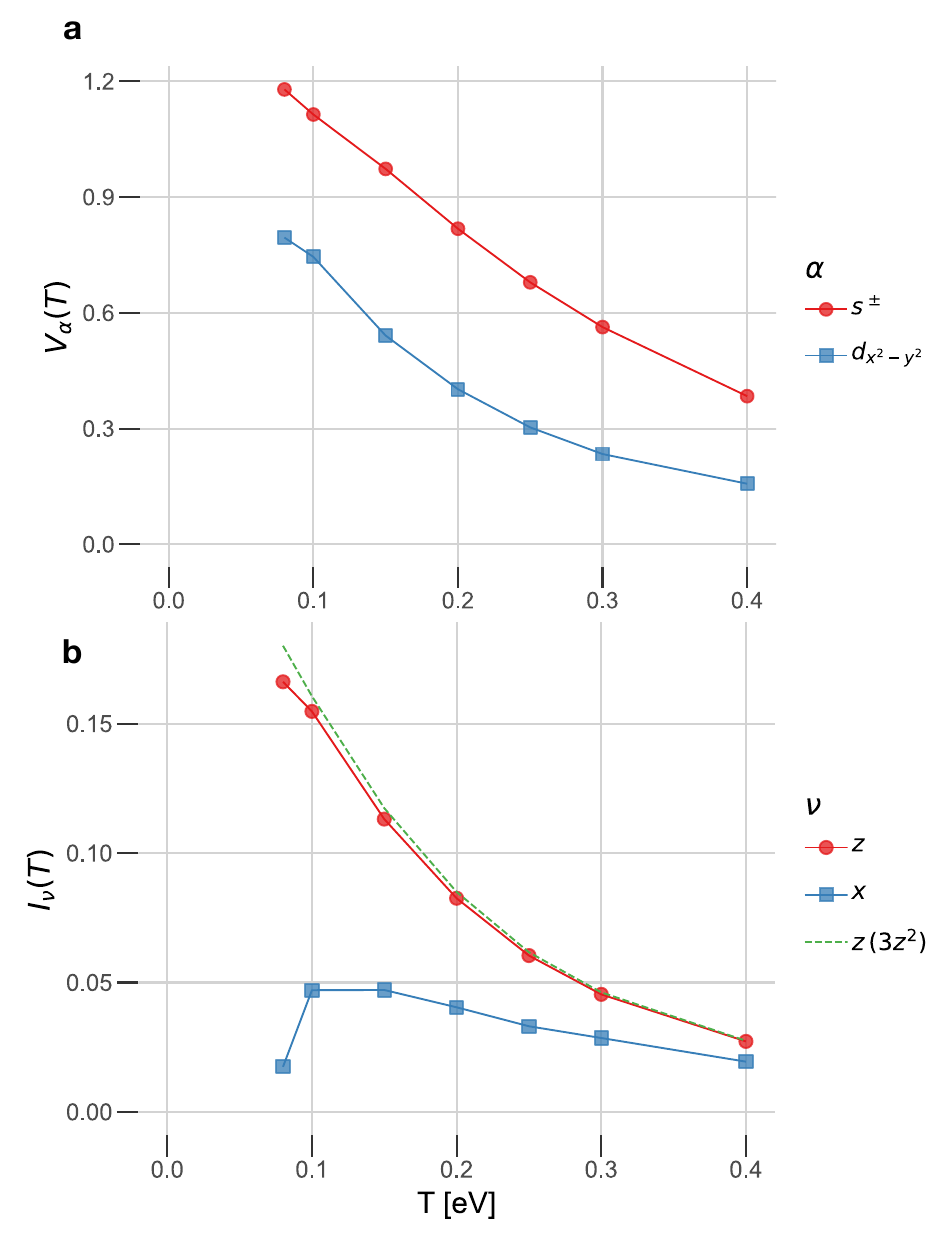}
  \caption{The strength of the pairing interaction $V_\alpha$ for $\alpha=s^\pm$ and $d_{x^2-y^2}$ calculated from Eq.~\ref{eq:valpha} (a) and the integrated spin fluctuation spectral weight $I_\nu$, Eq.~\ref{eq:inu}, versus temperature. The dashed line in panel (b) shows the $d_{3z^2-r^2}$ contribution to the $\nu=z$ spectral weight. Results are for the $N_c=16$ cluster and $U=3$~eV, $U'=2$~eV, and $J=0.5$~eV. \label{fig:5}}
\end{figure}

A measure of the strength of the pairing interaction giving rise to the pairing states $\alpha$ shown in Fig.~\ref{fig:3} is
\begin{equation} \label{eq:valpha}
    V_\alpha(T) = \frac{1}{P_{\phi^\alpha}^0(T)} - \frac{1}{P_{\phi^\alpha}(T)}\,.
\end{equation}
As detailed in the Supplemental Material, $P_{\phi^\alpha}$ is the pair-field susceptibility in Eq.~(\ref{eq:Palpha}) using for the form factor $g^{\ell\ell'}_\alpha=\phi^\alpha_{\ell\ell'}$, i.e., the leading eigenvectors shown in Fig.~\ref{fig:3}, and $P_{\phi^\alpha}^0$ is the leading (zeroth order) term in the Bethe-Salpeter expansion of $P_{\phi^\alpha}$. A plot of $V_\alpha(T)$ is shown in Fig.~\ref{fig:5}a. In Fig.~\ref{fig:5}b, we have plotted the integrated spin fluctuation spectral weights
\begin{align} \label{eq:inu}
    I_\nu & = \frac{1}{N_c} \sum_{\bm Q}\int \frac{d\omega}{\pi}\frac{\chi''(\bm Q, \omega)}{\omega} \cos Q_\nu \nonumber \\
          & = \frac{1}{N_c}\sum_{\bm Q}\chi(\bm Q) \cos Q_\nu\,
\end{align}
for the intra- ($Q_\nu=Q_x$) and inter-layer ($Q_\nu=Q_z$) near-neighbor spin fluctuations. Here, the sum over $\bm Q$ runs over the $N_c$ cluster momenta. The pairing interaction $V_\alpha(T)$ increases as the temperature $T$ is lowered for both pairing channels $\alpha=s^\pm$ and $d_{x^2-y^2}$. Similarly, the near-neighbor inter-layer spin-fluctuation spectral weight $I_z$ also increases as $T$ decreases. This increase arises from the increasing difference between the inter-layer antiferromagnetic ($q_z=\pi$) and ferromagnetic $(q_z=0)$ spin correlations shown in Fig.~\ref{fig:4} as $T$ decreases. In contrast, the intra-layer spectral weight $I_x$ remains small and even decreases at the lowest temperatures indicating that intra-layer spin fluctuations do not play an important role. The dashed line for the $\ell_1,\ell_2=d_{3z^2-r^2}$ contribution in panel (b) shows that $I_z$ primarily arises from spin fluctuations in the $d_{3z^2-r^2}$ orbital. Taken together, we believe that this demonstrates that \textit{both pairing states, $s^\pm$ and $d_{x^2-y^2}$, are primarily driven by inter-layer spin fluctuations arising from the $d_{3z^2-r^2}$ orbital.}

\section*{Discussion}

These results support the picture of a simple single-orbital bilayer Hubbard model first discussed by Sakakibara \textit{et al}. in Ref.~\cite{PhysRevLett.132.106002} in the context of weak-coupling perturbative calculations, where the $d_{3z^2-r^2}$ orbital is the relevant orbital and the $d_{x^2-y^2}$ orbital plays a secondary role. From Fig.~\ref{fig:3}b it is clear that for the leading $s^\pm$ state, pairing is mostly limited to the $d_{3z^2-r^2}$ orbital, and Fig.~\ref{fig:4} shows that it is the $d_{3z^2-r^2}$ inter-layer spin fluctuations that give rise to this state. This picture is reminiscent of the physics of a simpler single-obital bilayer Hubbard model with interlayer hopping $t_\perp$, for which DCA QMC calculations found a similar leading $s^\pm$ pairing state for parameter regimes where $t_\perp$ is larger than the in-plane nearest neighbor hopping $t$ \cite{maier_pair_2011}. The present two-orbital model has similar relative energy scales, where the inter-layer hopping $t^\perp_{zz}$ for the $d_{3z^2-r^2}$ orbital is the largest hopping in the model. Consistent with the present results, the earlier calculations for the single-orbital model also demonstrated that the inter-layer spin-fluctuations drive the leading $s^\pm$ state \cite{maier_pair_2011}. Our results provide non-perturbative support for this picture and strengthen the notion that $T_c$ in bilayer La$_3$Ni$_2$O$_7$ can be further increased because of its relation to the simple bilayer Hubbard model \cite{PhysRevLett.132.106002}, for which previous DCA QMC calculations found a significantly enhanced $T_c$ \cite{maier_pair_2011}. Strategies to achieve such an enhancement by tuning the electronic structure in ways that result in a self-doping of the $d_{3z^2-r^2}$ orbital were discussed in Ref.~\cite{PhysRevLett.132.106002}.

To conclude, in this work we have reported non-perturbative DCA QMC results for a realistic bilayer two-orbital Hubbard-Hund model of La$_3$Ni$_2$O$_7$ with physically relevant interaction parameters. We have shown that this model has an $s^\pm$ superconducting instability with singlet pairs formed primarily from electrons in the $d_{3z^2-r^2}$ orbitals on neighboring sites in the top and bottom layers, and that this pair formation is driven by strong inter-layer spin fluctuations arising from the $d_{3z^2-r^2}$ orbitals. These results support a simple bilayer Hubbard model picture of pairing in these compounds, for which similar calculations have found an enhanced $T_c$, supporting the possibility that $T_c$ in La$_3$Ni$_2$O$_7$ can be further increased through electronic structure tuning.


\section*{Methods}

The Hamiltonian of the bilayer two-orbital model we consider
\begin{equation}
H=H_0 + H_{\rm int}
\end{equation}
consists of a non-interacting tight-binding part $H_0$ and an interaction term $H_{\rm int}$. The non-interacting term is given by
\begin{align} \label{eq:H0}
H_0 &= \sum_{\langle ij \rangle}\sum_{\ell\ell', \sigma} t^{ij}_{\ell\ell'} (d^\dagger_{i\ell\sigma}d^{\phantom\dagger}_{j\ell'\sigma}+{\rm h.c.})+\epsilon_x \sum_{i\sigma} n_{ix\sigma}\nonumber\\
	&- \mu\sum_{i\ell\sigma} n_{i\ell\sigma}
\end{align}
Here, $t^{ij}_{\ell\ell'}$ is the hopping amplitude between orbitals $\ell,\,\ell' \in \{x\equiv d_{x^2-y^2},\, z\equiv d_{3z^2-r^2}\}$ on nearest neighbor sites $i,\, j$ in the bilayer lattice, $d^{\phantom\dagger}_{i\ell\sigma}$ ($d_{i\ell\sigma}^{\dagger}$) annihilates (creates) an electron with spin $\sigma$ on site $i$ in orbital $\ell$ and $n_{i\ell\sigma} = d^\dagger_{i\ell\sigma}d^{\phantom\dagger}_{i\ell\sigma}$. The values for $t^{ij}_{\ell\ell'}$ are given in the caption of Fig.~\ref{fig:1} and correspond to the system under 25~GPa pressure. We note that the largest hopping parameter is the intra-orbital, inter-layer hopping $t^\perp_{zz}$ between the $d_{3z^2-r^2}$ orbitals, leading to bonding and antibonding states as discussed in Ref.~\cite{zhang_structural_2024}. Finally, $\epsilon_x=0.506$~eV denotes the crystal field splitting between the $d_{x^2-y^2}$ and $d_{3z^2-r^2}$ orbitals. The chemical potential $\mu$ is adjusted so that the total site filling $\langle n \rangle = 1.5$.

The interaction term is given by
\begin{eqnarray}
H_{\rm int} &=& U \sum_{i} n_{i\ell\uparrow}n_{i\ell\downarrow} \nonumber \\
&+& \frac{U'}{2}\sum_i\sum_{\ell\neq\ell',\sigma,\sigma'} n_{i\ell\sigma} n_{i\ell'\sigma'} \nonumber \\
&+& \frac{U'-J}{2}\sum_i\sum_{\ell\neq\ell',\sigma} n_{i\ell\sigma} n_{i\ell'\sigma}
\end{eqnarray}
and contains the local intra-orbital and inter-orbital Coulomb repulsions $U$ and $U'$, respectively, as well as the density-density part of the Hund's rule coupling $J$. Here, we have neglected the spin-flip part of the Hund's rule coupling as well as the pair-hopping term. These terms are known to cause difficulties in the QMC treatment by significantly exacerbating the Fermion sign problem \cite{troyer_computational_2005}. These neglected terms have been shown to only mildly affect the phase diagram of a three-orbital Hubbard model \cite{PhysRevE.93.063313}. All the results presented in this paper are for $U=3$ eV, $U'=2$ eV, and $J=0.5$ eV. These values are representative of the parameters that have been determined from optical properties \cite{geisler_optical_2024} and have been used before in other calculations \cite{PhysRevLett.132.106002}.

We calculate the pairing and magnetic properties of this model using a DCA approximation \cite{hettler_nonlocal_1998,maier_quantum_2005}, which approximates the thermodynamic limit by a finite size $N_c$-site cluster that is self-consistently embedded in a dynamic mean-field designed to treat the effects of the remaining sites in the lattice. The effective cluster problem is solved with a continuous-time auxiliary field quantum Monte Carlo algorithm \cite{gull_continuous-time_2011,gull_submatrix_2011}. The efficient implementation of this DCA QMC algorithm in the DCA++ code \cite{hahner_dca_2020} allows us to perform sufficiently long sampling to acurately resolve the superconducting properties. We have used a two-site (1$\times$1)$\times$2 cluster ($N_c=2$) with one site in the bottom and one site in the top layer, an 8-site (2$\times$2)$\times$2 cluster with four (2$\times 2$) sites per layer ($N_c=8$), and a 16-site cluster with 8 sites (in a diamond shape) per layer ($N_c=16$).

\begin{acknowledgements}
This work was supported by the U.S. Department of Energy, Office of Science, Basic Energy Sciences, Materials Sciences and Engineering Division. This manuscript has been authored by UT-Battelle, LLC, under Contract No. DE-AC0500OR22725 with the U.S. Department of Energy.  An award of computer time was provided by the INCITE program. This research also used resources of the Oak Ridge Leadership Computing Facility, which is a DOE Office of Science User Facility supported under Contract DE-AC05-00OR22725. The United States Government retains and the publisher, by accepting the article for publication, acknowledges that the United States Government retains a nonexclusive, paid-up, irrevocable, world-wide license to publish or reproduce the published form of this manuscript, or allow others to do so, for the United States Government purposes. The Department of Energy will provide public access to these results of federally sponsored research in accordance with the DOE Public Access Plan (\url{http://energy.gov/downloads/doe-public-access-plan}).
\end{acknowledgements}

\bibliography{main}

\begin{thebibliography}{38}%
\makeatletter
\providecommand \@ifxundefined [1]{%
 \@ifx{#1\undefined}
}%
\providecommand \@ifnum [1]{%
 \ifnum #1\expandafter \@firstoftwo
 \else \expandafter \@secondoftwo
 \fi
}%
\providecommand \@ifx [1]{%
 \ifx #1\expandafter \@firstoftwo
 \else \expandafter \@secondoftwo
 \fi
}%
\providecommand \natexlab [1]{#1}%
\providecommand \enquote  [1]{``#1''}%
\providecommand \bibnamefont  [1]{#1}%
\providecommand \bibfnamefont [1]{#1}%
\providecommand \citenamefont [1]{#1}%
\providecommand \href@noop [0]{\@secondoftwo}%
\providecommand \href [0]{\begingroup \@sanitize@url \@href}%
\providecommand \@href[1]{\@@startlink{#1}\@@href}%
\providecommand \@@href[1]{\endgroup#1\@@endlink}%
\providecommand \@sanitize@url [0]{\catcode `\\12\catcode `\$12\catcode
  `\&12\catcode `\#12\catcode `\^12\catcode `\_12\catcode `\%12\relax}%
\providecommand \@@startlink[1]{}%
\providecommand \@@endlink[0]{}%
\providecommand \url  [0]{\begingroup\@sanitize@url \@url }%
\providecommand \@url [1]{\endgroup\@href {#1}{\urlprefix }}%
\providecommand \urlprefix  [0]{URL }%
\providecommand \Eprint [0]{\href }%
\providecommand \doibase [0]{https://doi.org/}%
\providecommand \selectlanguage [0]{\@gobble}%
\providecommand \bibinfo  [0]{\@secondoftwo}%
\providecommand \bibfield  [0]{\@secondoftwo}%
\providecommand \translation [1]{[#1]}%
\providecommand \BibitemOpen [0]{}%
\providecommand \bibitemStop [0]{}%
\providecommand \bibitemNoStop [0]{.\EOS\space}%
\providecommand \EOS [0]{\spacefactor3000\relax}%
\providecommand \BibitemShut  [1]{\csname bibitem#1\endcsname}%
\let\auto@bib@innerbib\@empty
\bibitem [{\citenamefont {Sun}\ \emph {et~al.}(2023)\citenamefont {Sun},
  \citenamefont {Huo}, \citenamefont {Hu}, \citenamefont {Li}, \citenamefont
  {Liu}, \citenamefont {Han}, \citenamefont {Tang}, \citenamefont {Mao},
  \citenamefont {Yang}, \citenamefont {Wang}, \citenamefont {Cheng},
  \citenamefont {Yao}, \citenamefont {Zhang},\ and\ \citenamefont
  {Wang}}]{sun_signatures_2023}%
  \BibitemOpen
  \bibfield  {author} {\bibinfo {author} {\bibfnamefont {H.}~\bibnamefont
  {Sun}}, \bibinfo {author} {\bibfnamefont {M.}~\bibnamefont {Huo}}, \bibinfo
  {author} {\bibfnamefont {X.}~\bibnamefont {Hu}}, \bibinfo {author}
  {\bibfnamefont {J.}~\bibnamefont {Li}}, \bibinfo {author} {\bibfnamefont
  {Z.}~\bibnamefont {Liu}}, \bibinfo {author} {\bibfnamefont {Y.}~\bibnamefont
  {Han}}, \bibinfo {author} {\bibfnamefont {L.}~\bibnamefont {Tang}}, \bibinfo
  {author} {\bibfnamefont {Z.}~\bibnamefont {Mao}}, \bibinfo {author}
  {\bibfnamefont {P.}~\bibnamefont {Yang}}, \bibinfo {author} {\bibfnamefont
  {B.}~\bibnamefont {Wang}}, \bibinfo {author} {\bibfnamefont {J.}~\bibnamefont
  {Cheng}}, \bibinfo {author} {\bibfnamefont {D.-X.}\ \bibnamefont {Yao}},
  \bibinfo {author} {\bibfnamefont {G.-M.}\ \bibnamefont {Zhang}},\ and\
  \bibinfo {author} {\bibfnamefont {M.}~\bibnamefont {Wang}},\ }\href
  {https://doi.org/10.1038/s41586-023-06408-7} {\bibfield  {journal} {\bibinfo
  {journal} {Nature}\ }\textbf {\bibinfo {volume} {621}},\ \bibinfo {pages}
  {493} (\bibinfo {year} {2023})}\BibitemShut {NoStop}%
\bibitem [{\citenamefont {Hou}\ \emph {et~al.}(2023)\citenamefont {Hou},
  \citenamefont {Yang}, \citenamefont {Liu}, \citenamefont {Li}, \citenamefont
  {Shan}, \citenamefont {Ma}, \citenamefont {Wang}, \citenamefont {Wang},
  \citenamefont {Guo}, \citenamefont {Sun}, \citenamefont {Uwatoko},
  \citenamefont {Wang}, \citenamefont {Zhang}, \citenamefont {Wang},\ and\
  \citenamefont {Cheng}}]{hou_emergence_2023}%
  \BibitemOpen
  \bibfield  {author} {\bibinfo {author} {\bibfnamefont {J.}~\bibnamefont
  {Hou}}, \bibinfo {author} {\bibfnamefont {P.-T.}\ \bibnamefont {Yang}},
  \bibinfo {author} {\bibfnamefont {Z.-Y.}\ \bibnamefont {Liu}}, \bibinfo
  {author} {\bibfnamefont {J.-Y.}\ \bibnamefont {Li}}, \bibinfo {author}
  {\bibfnamefont {P.-F.}\ \bibnamefont {Shan}}, \bibinfo {author}
  {\bibfnamefont {L.}~\bibnamefont {Ma}}, \bibinfo {author} {\bibfnamefont
  {G.}~\bibnamefont {Wang}}, \bibinfo {author} {\bibfnamefont {N.-N.}\
  \bibnamefont {Wang}}, \bibinfo {author} {\bibfnamefont {H.-Z.}\ \bibnamefont
  {Guo}}, \bibinfo {author} {\bibfnamefont {J.-P.}\ \bibnamefont {Sun}},
  \bibinfo {author} {\bibfnamefont {Y.}~\bibnamefont {Uwatoko}}, \bibinfo
  {author} {\bibfnamefont {M.}~\bibnamefont {Wang}}, \bibinfo {author}
  {\bibfnamefont {G.-M.}\ \bibnamefont {Zhang}}, \bibinfo {author}
  {\bibfnamefont {B.-S.}\ \bibnamefont {Wang}},\ and\ \bibinfo {author}
  {\bibfnamefont {J.-G.}\ \bibnamefont {Cheng}},\ }\bibfield  {journal}
  {\bibinfo  {journal} {Chin. Phys. Lett.}\ }\textbf {\bibinfo {volume} {40}},\
  \href {https://doi.org/10.1088/0256-307X/40/11/117302}
  {10.1088/0256-307X/40/11/117302} (\bibinfo {year} {2023})\BibitemShut
  {NoStop}%
\bibitem [{\citenamefont {Luo}\ \emph {et~al.}(2023)\citenamefont {Luo},
  \citenamefont {Hu}, \citenamefont {Wang}, \citenamefont {W\'u},\ and\
  \citenamefont {Yao}}]{PhysRevLett.131.126001}%
  \BibitemOpen
  \bibfield  {author} {\bibinfo {author} {\bibfnamefont {Z.}~\bibnamefont
  {Luo}}, \bibinfo {author} {\bibfnamefont {X.}~\bibnamefont {Hu}}, \bibinfo
  {author} {\bibfnamefont {M.}~\bibnamefont {Wang}}, \bibinfo {author}
  {\bibfnamefont {W.}~\bibnamefont {W\'u}},\ and\ \bibinfo {author}
  {\bibfnamefont {D.-X.}\ \bibnamefont {Yao}},\ }\href
  {https://doi.org/10.1103/PhysRevLett.131.126001} {\bibfield  {journal}
  {\bibinfo  {journal} {Phys. Rev. Lett.}\ }\textbf {\bibinfo {volume} {131}},\
  \bibinfo {pages} {126001} (\bibinfo {year} {2023})}\BibitemShut {NoStop}%
\bibitem [{\citenamefont {Liu}\ \emph {et~al.}(2023)\citenamefont {Liu},
  \citenamefont {Mei}, \citenamefont {Ye}, \citenamefont {Chen},\ and\
  \citenamefont {Yang}}]{PhysRevLett.131.236002}%
  \BibitemOpen
  \bibfield  {author} {\bibinfo {author} {\bibfnamefont {Y.-B.}\ \bibnamefont
  {Liu}}, \bibinfo {author} {\bibfnamefont {J.-W.}\ \bibnamefont {Mei}},
  \bibinfo {author} {\bibfnamefont {F.}~\bibnamefont {Ye}}, \bibinfo {author}
  {\bibfnamefont {W.-Q.}\ \bibnamefont {Chen}},\ and\ \bibinfo {author}
  {\bibfnamefont {F.}~\bibnamefont {Yang}},\ }\href
  {https://doi.org/10.1103/PhysRevLett.131.236002} {\bibfield  {journal}
  {\bibinfo  {journal} {Phys. Rev. Lett.}\ }\textbf {\bibinfo {volume} {131}},\
  \bibinfo {pages} {236002} (\bibinfo {year} {2023})}\BibitemShut {NoStop}%
\bibitem [{\citenamefont {Lechermann}\ \emph {et~al.}(2023)\citenamefont
  {Lechermann}, \citenamefont {Gondolf}, \citenamefont {B\"otzel},\ and\
  \citenamefont {Eremin}}]{PhysRevB.108.L201121}%
  \BibitemOpen
  \bibfield  {author} {\bibinfo {author} {\bibfnamefont {F.}~\bibnamefont
  {Lechermann}}, \bibinfo {author} {\bibfnamefont {J.}~\bibnamefont {Gondolf}},
  \bibinfo {author} {\bibfnamefont {S.}~\bibnamefont {B\"otzel}},\ and\
  \bibinfo {author} {\bibfnamefont {I.~M.}\ \bibnamefont {Eremin}},\ }\href
  {https://doi.org/10.1103/PhysRevB.108.L201121} {\bibfield  {journal}
  {\bibinfo  {journal} {Phys. Rev. B}\ }\textbf {\bibinfo {volume} {108}},\
  \bibinfo {pages} {L201121} (\bibinfo {year} {2023})}\BibitemShut {NoStop}%
\bibitem [{\citenamefont {Zhang}\ \emph {et~al.}(2024)\citenamefont {Zhang},
  \citenamefont {Lin}, \citenamefont {Moreo}, \citenamefont {Maier},\ and\
  \citenamefont {Dagotto}}]{zhang_structural_2024}%
  \BibitemOpen
  \bibfield  {author} {\bibinfo {author} {\bibfnamefont {Y.}~\bibnamefont
  {Zhang}}, \bibinfo {author} {\bibfnamefont {L.-F.}\ \bibnamefont {Lin}},
  \bibinfo {author} {\bibfnamefont {A.}~\bibnamefont {Moreo}}, \bibinfo
  {author} {\bibfnamefont {T.~A.}\ \bibnamefont {Maier}},\ and\ \bibinfo
  {author} {\bibfnamefont {E.}~\bibnamefont {Dagotto}},\ }\href
  {https://doi.org/10.1038/s41467-024-46622-z} {\bibfield  {journal} {\bibinfo
  {journal} {Nat. Comm.}\ }\textbf {\bibinfo {volume} {15}},\ \bibinfo {pages}
  {2470} (\bibinfo {year} {2024})}\BibitemShut {NoStop}%
\bibitem [{\citenamefont {Sakakibara}\ \emph {et~al.}(2024)\citenamefont
  {Sakakibara}, \citenamefont {Kitamine}, \citenamefont {Ochi},\ and\
  \citenamefont {Kuroki}}]{PhysRevLett.132.106002}%
  \BibitemOpen
  \bibfield  {author} {\bibinfo {author} {\bibfnamefont {H.}~\bibnamefont
  {Sakakibara}}, \bibinfo {author} {\bibfnamefont {N.}~\bibnamefont
  {Kitamine}}, \bibinfo {author} {\bibfnamefont {M.}~\bibnamefont {Ochi}},\
  and\ \bibinfo {author} {\bibfnamefont {K.}~\bibnamefont {Kuroki}},\ }\href
  {https://doi.org/10.1103/PhysRevLett.132.106002} {\bibfield  {journal}
  {\bibinfo  {journal} {Phys. Rev. Lett.}\ }\textbf {\bibinfo {volume} {132}},\
  \bibinfo {pages} {106002} (\bibinfo {year} {2024})}\BibitemShut {NoStop}%
\bibitem [{\citenamefont {Gu}\ \emph {et~al.}(2025)\citenamefont {Gu},
  \citenamefont {Le}, \citenamefont {Yang}, \citenamefont {Wu},\ and\
  \citenamefont {Hu}}]{gu_effective_2025}%
  \BibitemOpen
  \bibfield  {author} {\bibinfo {author} {\bibfnamefont {Y.}~\bibnamefont
  {Gu}}, \bibinfo {author} {\bibfnamefont {C.}~\bibnamefont {Le}}, \bibinfo
  {author} {\bibfnamefont {Z.}~\bibnamefont {Yang}}, \bibinfo {author}
  {\bibfnamefont {X.}~\bibnamefont {Wu}},\ and\ \bibinfo {author}
  {\bibfnamefont {J.}~\bibnamefont {Hu}},\ }\href
  {https://doi.org/10.1103/physrevb.111.174506} {\bibfield  {journal} {\bibinfo
   {journal} {Phys. Rev. B}\ }\textbf {\bibinfo {volume} {111}},\ \bibinfo
  {pages} {174506} (\bibinfo {year} {2025})}\BibitemShut {NoStop}%
\bibitem [{\citenamefont {Xi}\ \emph {et~al.}(2025)\citenamefont {Xi},
  \citenamefont {Yu},\ and\ \citenamefont {Li}}]{xi_transition_2025}%
  \BibitemOpen
  \bibfield  {author} {\bibinfo {author} {\bibfnamefont {W.}~\bibnamefont
  {Xi}}, \bibinfo {author} {\bibfnamefont {S.-L.}\ \bibnamefont {Yu}},\ and\
  \bibinfo {author} {\bibfnamefont {J.-X.}\ \bibnamefont {Li}},\ }\href
  {https://doi.org/10.1103/physrevb.111.104505} {\bibfield  {journal} {\bibinfo
   {journal} {Phys. Rev. B}\ }\textbf {\bibinfo {volume} {111}},\ \bibinfo
  {pages} {104505} (\bibinfo {year} {2025})}\BibitemShut {NoStop}%
\bibitem [{\citenamefont {Yang}\ \emph
  {et~al.}(2023{\natexlab{a}})\citenamefont {Yang}, \citenamefont {Zhang},\
  and\ \citenamefont {Zhang}}]{YangYf_2023_Interlayer}%
  \BibitemOpen
  \bibfield  {author} {\bibinfo {author} {\bibfnamefont {Y.-f.}\ \bibnamefont
  {Yang}}, \bibinfo {author} {\bibfnamefont {G.-M.}\ \bibnamefont {Zhang}},\
  and\ \bibinfo {author} {\bibfnamefont {F.-C.}\ \bibnamefont {Zhang}},\ }\href
  {https://doi.org/10.1103/PhysRevB.108.L201108} {\bibfield  {journal}
  {\bibinfo  {journal} {Phys. Rev. B}\ }\textbf {\bibinfo {volume} {108}},\
  \bibinfo {pages} {L201108} (\bibinfo {year}
  {2023}{\natexlab{a}})}\BibitemShut {NoStop}%
\bibitem [{\citenamefont {B\"otzel}\ \emph {et~al.}(2024)\citenamefont
  {B\"otzel}, \citenamefont {Lechermann}, \citenamefont {Gondolf},\ and\
  \citenamefont {Eremin}}]{BotzelS_2024_Theory}%
  \BibitemOpen
  \bibfield  {author} {\bibinfo {author} {\bibfnamefont {S.}~\bibnamefont
  {B\"otzel}}, \bibinfo {author} {\bibfnamefont {F.}~\bibnamefont
  {Lechermann}}, \bibinfo {author} {\bibfnamefont {J.}~\bibnamefont
  {Gondolf}},\ and\ \bibinfo {author} {\bibfnamefont {I.~M.}\ \bibnamefont
  {Eremin}},\ }\href {https://doi.org/10.1103/PhysRevB.109.L180502} {\bibfield
  {journal} {\bibinfo  {journal} {Phys. Rev. B}\ }\textbf {\bibinfo {volume}
  {109}},\ \bibinfo {pages} {L180502} (\bibinfo {year} {2024})}\BibitemShut
  {NoStop}%
\bibitem [{\citenamefont {Xia}\ \emph {et~al.}(2025)\citenamefont {Xia},
  \citenamefont {Liu}, \citenamefont {Zhou},\ and\ \citenamefont
  {Chen}}]{xia_sensitive_2025}%
  \BibitemOpen
  \bibfield  {author} {\bibinfo {author} {\bibfnamefont {C.}~\bibnamefont
  {Xia}}, \bibinfo {author} {\bibfnamefont {H.}~\bibnamefont {Liu}}, \bibinfo
  {author} {\bibfnamefont {S.}~\bibnamefont {Zhou}},\ and\ \bibinfo {author}
  {\bibfnamefont {H.}~\bibnamefont {Chen}},\ }\href
  {https://doi.org/10.1038/s41467-025-56206-0} {\bibfield  {journal} {\bibinfo
  {journal} {Nat. Comm.}\ }\textbf {\bibinfo {volume} {16}},\ \bibinfo {pages}
  {1054} (\bibinfo {year} {2025})}\BibitemShut {NoStop}%
\bibitem [{\citenamefont {Heier}\ \emph {et~al.}(2024)\citenamefont {Heier},
  \citenamefont {Park},\ and\ \citenamefont
  {Savrasov}}]{Heier_G_2024_Competing}%
  \BibitemOpen
  \bibfield  {author} {\bibinfo {author} {\bibfnamefont {G.}~\bibnamefont
  {Heier}}, \bibinfo {author} {\bibfnamefont {K.}~\bibnamefont {Park}},\ and\
  \bibinfo {author} {\bibfnamefont {S.~Y.}\ \bibnamefont {Savrasov}},\ }\href
  {https://doi.org/10.1103/PhysRevB.109.104508} {\bibfield  {journal} {\bibinfo
   {journal} {Phys. Rev. B}\ }\textbf {\bibinfo {volume} {109}},\ \bibinfo
  {pages} {104508} (\bibinfo {year} {2024})}\BibitemShut {NoStop}%
\bibitem [{\citenamefont {Yang}\ \emph
  {et~al.}(2023{\natexlab{b}})\citenamefont {Yang}, \citenamefont {Wang},\ and\
  \citenamefont {Wang}}]{YangQG_2023_Possible}%
  \BibitemOpen
  \bibfield  {author} {\bibinfo {author} {\bibfnamefont {Q.-G.}\ \bibnamefont
  {Yang}}, \bibinfo {author} {\bibfnamefont {D.}~\bibnamefont {Wang}},\ and\
  \bibinfo {author} {\bibfnamefont {Q.-H.}\ \bibnamefont {Wang}},\ }\href
  {https://doi.org/10.1103/PhysRevB.108.L140505} {\bibfield  {journal}
  {\bibinfo  {journal} {Phys. Rev. B}\ }\textbf {\bibinfo {volume} {108}},\
  \bibinfo {pages} {L140505} (\bibinfo {year}
  {2023}{\natexlab{b}})}\BibitemShut {NoStop}%
\bibitem [{\citenamefont {Lu}\ \emph {et~al.}(2024)\citenamefont {Lu},
  \citenamefont {Pan}, \citenamefont {Yang},\ and\ \citenamefont
  {Wu}}]{Lu_C_2024_Interlayer}%
  \BibitemOpen
  \bibfield  {author} {\bibinfo {author} {\bibfnamefont {C.}~\bibnamefont
  {Lu}}, \bibinfo {author} {\bibfnamefont {Z.}~\bibnamefont {Pan}}, \bibinfo
  {author} {\bibfnamefont {F.}~\bibnamefont {Yang}},\ and\ \bibinfo {author}
  {\bibfnamefont {C.}~\bibnamefont {Wu}},\ }\href
  {https://doi.org/10.1103/PhysRevLett.132.146002} {\bibfield  {journal}
  {\bibinfo  {journal} {Phys. Rev. Lett.}\ }\textbf {\bibinfo {volume} {132}},\
  \bibinfo {pages} {146002} (\bibinfo {year} {2024})}\BibitemShut {NoStop}%
\bibitem [{\citenamefont {Liao}\ \emph {et~al.}(2023)\citenamefont {Liao},
  \citenamefont {Chen}, \citenamefont {Duan}, \citenamefont {Wang},
  \citenamefont {Liu}, \citenamefont {Yu},\ and\ \citenamefont
  {Si}}]{liaoZ_2023_Electron}%
  \BibitemOpen
  \bibfield  {author} {\bibinfo {author} {\bibfnamefont {Z.}~\bibnamefont
  {Liao}}, \bibinfo {author} {\bibfnamefont {L.}~\bibnamefont {Chen}}, \bibinfo
  {author} {\bibfnamefont {G.}~\bibnamefont {Duan}}, \bibinfo {author}
  {\bibfnamefont {Y.}~\bibnamefont {Wang}}, \bibinfo {author} {\bibfnamefont
  {C.}~\bibnamefont {Liu}}, \bibinfo {author} {\bibfnamefont {R.}~\bibnamefont
  {Yu}},\ and\ \bibinfo {author} {\bibfnamefont {Q.}~\bibnamefont {Si}},\
  }\href {https://doi.org/10.1103/PhysRevB.108.214522} {\bibfield  {journal}
  {\bibinfo  {journal} {Phys. Rev. B}\ }\textbf {\bibinfo {volume} {108}},\
  \bibinfo {pages} {214522} (\bibinfo {year} {2023})}\BibitemShut {NoStop}%
\bibitem [{\citenamefont {Jiang}\ \emph {et~al.}(2024)\citenamefont {Jiang},
  \citenamefont {Wang},\ and\ \citenamefont {Zhang}}]{Jiang_2024}%
  \BibitemOpen
  \bibfield  {author} {\bibinfo {author} {\bibfnamefont {K.}~\bibnamefont
  {Jiang}}, \bibinfo {author} {\bibfnamefont {Z.}~\bibnamefont {Wang}},\ and\
  \bibinfo {author} {\bibfnamefont {F.-C.}\ \bibnamefont {Zhang}},\ }\href
  {https://doi.org/10.1088/0256-307X/41/1/017402} {\bibfield  {journal}
  {\bibinfo  {journal} {Chin. Phys. Lett.}\ }\textbf {\bibinfo {volume} {41}},\
  \bibinfo {pages} {017402} (\bibinfo {year} {2024})}\BibitemShut {NoStop}%
\bibitem [{\citenamefont {Liu}\ \emph {et~al.}(2024)\citenamefont {Liu},
  \citenamefont {Huo}, \citenamefont {Li}, \citenamefont {Li}, \citenamefont
  {Liu}, \citenamefont {Dai}, \citenamefont {Zhou}, \citenamefont {Hao},
  \citenamefont {Lu}, \citenamefont {Wang},\ and\ \citenamefont
  {Wen}}]{liu_electronic_2024}%
  \BibitemOpen
  \bibfield  {author} {\bibinfo {author} {\bibfnamefont {Z.}~\bibnamefont
  {Liu}}, \bibinfo {author} {\bibfnamefont {M.}~\bibnamefont {Huo}}, \bibinfo
  {author} {\bibfnamefont {J.}~\bibnamefont {Li}}, \bibinfo {author}
  {\bibfnamefont {Q.}~\bibnamefont {Li}}, \bibinfo {author} {\bibfnamefont
  {Y.}~\bibnamefont {Liu}}, \bibinfo {author} {\bibfnamefont {Y.}~\bibnamefont
  {Dai}}, \bibinfo {author} {\bibfnamefont {X.}~\bibnamefont {Zhou}}, \bibinfo
  {author} {\bibfnamefont {J.}~\bibnamefont {Hao}}, \bibinfo {author}
  {\bibfnamefont {Y.}~\bibnamefont {Lu}}, \bibinfo {author} {\bibfnamefont
  {M.}~\bibnamefont {Wang}},\ and\ \bibinfo {author} {\bibfnamefont {H.-H.}\
  \bibnamefont {Wen}},\ }\href {https://doi.org/10.1038/s41467-024-52001-5}
  {\bibfield  {journal} {\bibinfo  {journal} {Nat. Comm.}\ }\textbf {\bibinfo
  {volume} {15}},\ \bibinfo {pages} {7570} (\bibinfo {year}
  {2024})}\BibitemShut {NoStop}%
\bibitem [{\citenamefont {Geisler}\ \emph {et~al.}(2024)\citenamefont
  {Geisler}, \citenamefont {Fanfarillo}, \citenamefont {Hamlin}, \citenamefont
  {Stewart}, \citenamefont {Hennig},\ and\ \citenamefont
  {Hirschfeld}}]{geisler_optical_2024}%
  \BibitemOpen
  \bibfield  {author} {\bibinfo {author} {\bibfnamefont {B.}~\bibnamefont
  {Geisler}}, \bibinfo {author} {\bibfnamefont {L.}~\bibnamefont {Fanfarillo}},
  \bibinfo {author} {\bibfnamefont {J.~J.}\ \bibnamefont {Hamlin}}, \bibinfo
  {author} {\bibfnamefont {G.~R.}\ \bibnamefont {Stewart}}, \bibinfo {author}
  {\bibfnamefont {R.~G.}\ \bibnamefont {Hennig}},\ and\ \bibinfo {author}
  {\bibfnamefont {P.~J.}\ \bibnamefont {Hirschfeld}},\ }\href
  {https://doi.org/10.1038/s41535-024-00690-y} {\bibfield  {journal} {\bibinfo
  {journal} {npj Quantum Materials}\ }\textbf {\bibinfo {volume} {9}},\
  \bibinfo {pages} {89} (\bibinfo {year} {2024})}\BibitemShut {NoStop}%
\bibitem [{\citenamefont {Yue}\ \emph {et~al.}(2025)\citenamefont {Yue},
  \citenamefont {Miao}, \citenamefont {Huang}, \citenamefont {Hua},
  \citenamefont {Li}, \citenamefont {Li}, \citenamefont {Zhou}, \citenamefont
  {Lv}, \citenamefont {Yang}, \citenamefont {Sun}, \citenamefont {Sun},
  \citenamefont {Lin}, \citenamefont {Xue}, \citenamefont {Chen},\ and\
  \citenamefont {Chen}}]{yue2025correlatedelectronicstructuresunconventional}%
  \BibitemOpen
  \bibfield  {author} {\bibinfo {author} {\bibfnamefont {C.}~\bibnamefont
  {Yue}}, \bibinfo {author} {\bibfnamefont {J.-J.}\ \bibnamefont {Miao}},
  \bibinfo {author} {\bibfnamefont {H.}~\bibnamefont {Huang}}, \bibinfo
  {author} {\bibfnamefont {Y.}~\bibnamefont {Hua}}, \bibinfo {author}
  {\bibfnamefont {P.}~\bibnamefont {Li}}, \bibinfo {author} {\bibfnamefont
  {Y.}~\bibnamefont {Li}}, \bibinfo {author} {\bibfnamefont {G.}~\bibnamefont
  {Zhou}}, \bibinfo {author} {\bibfnamefont {W.}~\bibnamefont {Lv}}, \bibinfo
  {author} {\bibfnamefont {Q.}~\bibnamefont {Yang}}, \bibinfo {author}
  {\bibfnamefont {H.}~\bibnamefont {Sun}}, \bibinfo {author} {\bibfnamefont
  {Y.-J.}\ \bibnamefont {Sun}}, \bibinfo {author} {\bibfnamefont
  {J.}~\bibnamefont {Lin}}, \bibinfo {author} {\bibfnamefont {Q.-K.}\
  \bibnamefont {Xue}}, \bibinfo {author} {\bibfnamefont {Z.}~\bibnamefont
  {Chen}},\ and\ \bibinfo {author} {\bibfnamefont {W.-Q.}\ \bibnamefont
  {Chen}},\ }\href {https://arxiv.org/abs/2501.06875} {\bibinfo {title}
  {Correlated electronic structures and unconventional superconductivity in
  bilayer nickelate heterostructures}} (\bibinfo {year} {2025}),\ \Eprint
  {https://arxiv.org/abs/2501.06875} {arXiv:2501.06875 [cond-mat.str-el]}
  \BibitemShut {NoStop}%
\bibitem [{\citenamefont {Shen}\ \emph {et~al.}(2023)\citenamefont {Shen},
  \citenamefont {Qin},\ and\ \citenamefont {Zhang}}]{Shen_2023}%
  \BibitemOpen
  \bibfield  {author} {\bibinfo {author} {\bibfnamefont {Y.}~\bibnamefont
  {Shen}}, \bibinfo {author} {\bibfnamefont {M.}~\bibnamefont {Qin}},\ and\
  \bibinfo {author} {\bibfnamefont {G.-M.}\ \bibnamefont {Zhang}},\ }\href
  {https://doi.org/10.1088/0256-307X/40/12/127401} {\bibfield  {journal}
  {\bibinfo  {journal} {Chin. Phys. Lett.}\ }\textbf {\bibinfo {volume} {40}},\
  \bibinfo {pages} {127401} (\bibinfo {year} {2023})}\BibitemShut {NoStop}%
\bibitem [{\citenamefont {Qu}\ \emph {et~al.}(2024)\citenamefont {Qu},
  \citenamefont {Qu}, \citenamefont {Chen}, \citenamefont {Wu}, \citenamefont
  {Yang}, \citenamefont {Li},\ and\ \citenamefont {Su}}]{QuXZ_2024_Bilayer}%
  \BibitemOpen
  \bibfield  {author} {\bibinfo {author} {\bibfnamefont {X.-Z.}\ \bibnamefont
  {Qu}}, \bibinfo {author} {\bibfnamefont {D.-W.}\ \bibnamefont {Qu}}, \bibinfo
  {author} {\bibfnamefont {J.}~\bibnamefont {Chen}}, \bibinfo {author}
  {\bibfnamefont {C.}~\bibnamefont {Wu}}, \bibinfo {author} {\bibfnamefont
  {F.}~\bibnamefont {Yang}}, \bibinfo {author} {\bibfnamefont {W.}~\bibnamefont
  {Li}},\ and\ \bibinfo {author} {\bibfnamefont {G.}~\bibnamefont {Su}},\
  }\href {https://doi.org/10.1103/PhysRevLett.132.036502} {\bibfield  {journal}
  {\bibinfo  {journal} {Phys. Rev. Lett.}\ }\textbf {\bibinfo {volume} {132}},\
  \bibinfo {pages} {036502} (\bibinfo {year} {2024})}\BibitemShut {NoStop}%
\bibitem [{\citenamefont {Schlömer}\ \emph {et~al.}(2024)\citenamefont
  {Schlömer}, \citenamefont {Schollwöck}, \citenamefont {Grusdt},\ and\
  \citenamefont {Bohrdt}}]{schlomer_superconductivity_2024}%
  \BibitemOpen
  \bibfield  {author} {\bibinfo {author} {\bibfnamefont {H.}~\bibnamefont
  {Schlömer}}, \bibinfo {author} {\bibfnamefont {U.}~\bibnamefont
  {Schollwöck}}, \bibinfo {author} {\bibfnamefont {F.}~\bibnamefont
  {Grusdt}},\ and\ \bibinfo {author} {\bibfnamefont {A.}~\bibnamefont
  {Bohrdt}},\ }\href {https://doi.org/10.1038/s42005-024-01854-9} {\bibfield
  {journal} {\bibinfo  {journal} {Comm. Phys.}\ }\textbf {\bibinfo {volume}
  {7}},\ \bibinfo {pages} {366} (\bibinfo {year} {2024})}\BibitemShut {NoStop}%
\bibitem [{\citenamefont {Kaneko}\ \emph {et~al.}(2024)\citenamefont {Kaneko},
  \citenamefont {Sakakibara}, \citenamefont {Ochi},\ and\ \citenamefont
  {Kuroki}}]{PhysRevB.109.045154}%
  \BibitemOpen
  \bibfield  {author} {\bibinfo {author} {\bibfnamefont {T.}~\bibnamefont
  {Kaneko}}, \bibinfo {author} {\bibfnamefont {H.}~\bibnamefont {Sakakibara}},
  \bibinfo {author} {\bibfnamefont {M.}~\bibnamefont {Ochi}},\ and\ \bibinfo
  {author} {\bibfnamefont {K.}~\bibnamefont {Kuroki}},\ }\href
  {https://doi.org/10.1103/PhysRevB.109.045154} {\bibfield  {journal} {\bibinfo
   {journal} {Phys. Rev. B}\ }\textbf {\bibinfo {volume} {109}},\ \bibinfo
  {pages} {045154} (\bibinfo {year} {2024})}\BibitemShut {NoStop}%
\bibitem [{\citenamefont {Kakoi}\ \emph {et~al.}(2024)\citenamefont {Kakoi},
  \citenamefont {Kaneko}, \citenamefont {Sakakibara}, \citenamefont {Ochi},\
  and\ \citenamefont {Kuroki}}]{PhysRevB.109.L201124}%
  \BibitemOpen
  \bibfield  {author} {\bibinfo {author} {\bibfnamefont {M.}~\bibnamefont
  {Kakoi}}, \bibinfo {author} {\bibfnamefont {T.}~\bibnamefont {Kaneko}},
  \bibinfo {author} {\bibfnamefont {H.}~\bibnamefont {Sakakibara}}, \bibinfo
  {author} {\bibfnamefont {M.}~\bibnamefont {Ochi}},\ and\ \bibinfo {author}
  {\bibfnamefont {K.}~\bibnamefont {Kuroki}},\ }\href
  {https://doi.org/10.1103/PhysRevB.109.L201124} {\bibfield  {journal}
  {\bibinfo  {journal} {Phys. Rev. B}\ }\textbf {\bibinfo {volume} {109}},\
  \bibinfo {pages} {L201124} (\bibinfo {year} {2024})}\BibitemShut {NoStop}%
\bibitem [{\citenamefont {Qin}\ and\ \citenamefont
  {Yang}(2023)}]{PhysRevB.108.L140504}%
  \BibitemOpen
  \bibfield  {author} {\bibinfo {author} {\bibfnamefont {Q.}~\bibnamefont
  {Qin}}\ and\ \bibinfo {author} {\bibfnamefont {Y.-f.}\ \bibnamefont {Yang}},\
  }\href {https://doi.org/10.1103/PhysRevB.108.L140504} {\bibfield  {journal}
  {\bibinfo  {journal} {Phys. Rev. B}\ }\textbf {\bibinfo {volume} {108}},\
  \bibinfo {pages} {L140504} (\bibinfo {year} {2023})}\BibitemShut {NoStop}%
\bibitem [{\citenamefont {Zheng}\ and\ \citenamefont
  {Wú}(2025)}]{zheng_s_ifmmodepmelsetextpmfi-wave_2025}%
  \BibitemOpen
  \bibfield  {author} {\bibinfo {author} {\bibfnamefont {Y.-Y.}\ \bibnamefont
  {Zheng}}\ and\ \bibinfo {author} {\bibfnamefont {W.}~\bibnamefont {Wú}},\
  }\href {https://doi.org/10.1103/PhysRevB.111.035108} {\bibfield  {journal}
  {\bibinfo  {journal} {Phys. Rev. B}\ }\textbf {\bibinfo {volume} {111}},\
  \bibinfo {pages} {035108} (\bibinfo {year} {2025})}\BibitemShut {NoStop}%
\bibitem [{\citenamefont {Tian}\ \emph {et~al.}(2024)\citenamefont {Tian},
  \citenamefont {Chen}, \citenamefont {Wang}, \citenamefont {He},\ and\
  \citenamefont {Lu}}]{tian_correlation_2024}%
  \BibitemOpen
  \bibfield  {author} {\bibinfo {author} {\bibfnamefont {Y.-H.}\ \bibnamefont
  {Tian}}, \bibinfo {author} {\bibfnamefont {Y.}~\bibnamefont {Chen}}, \bibinfo
  {author} {\bibfnamefont {J.-M.}\ \bibnamefont {Wang}}, \bibinfo {author}
  {\bibfnamefont {R.-Q.}\ \bibnamefont {He}},\ and\ \bibinfo {author}
  {\bibfnamefont {Z.-Y.}\ \bibnamefont {Lu}},\ }\href
  {https://doi.org/10.1103/PhysRevB.109.165154} {\bibfield  {journal} {\bibinfo
   {journal} {Phys. Rev. B}\ }\textbf {\bibinfo {volume} {109}},\ \bibinfo
  {pages} {165154} (\bibinfo {year} {2024})}\BibitemShut {NoStop}%
\bibitem [{\citenamefont {Maier}\ \emph
  {et~al.}(2005{\natexlab{a}})\citenamefont {Maier}, \citenamefont {Jarrell},
  \citenamefont {Pruschke},\ and\ \citenamefont
  {Hettler}}]{maier_quantum_2005}%
  \BibitemOpen
  \bibfield  {author} {\bibinfo {author} {\bibfnamefont {T.}~\bibnamefont
  {Maier}}, \bibinfo {author} {\bibfnamefont {M.}~\bibnamefont {Jarrell}},
  \bibinfo {author} {\bibfnamefont {T.}~\bibnamefont {Pruschke}},\ and\
  \bibinfo {author} {\bibfnamefont {M.}~\bibnamefont {Hettler}},\ }\href
  {http://publish.aps.org/} {\bibfield  {journal} {\bibinfo  {journal} {Rev.
  Mod. Phys.}\ }\textbf {\bibinfo {volume} {77}},\ \bibinfo {pages} {1027}
  (\bibinfo {year} {2005}{\natexlab{a}})}\BibitemShut {NoStop}%
\bibitem [{\citenamefont {Maier}\ and\ \citenamefont
  {Scalapino}(2011)}]{maier_pair_2011}%
  \BibitemOpen
  \bibfield  {author} {\bibinfo {author} {\bibfnamefont {T.~A.}\ \bibnamefont
  {Maier}}\ and\ \bibinfo {author} {\bibfnamefont {D.~J.}\ \bibnamefont
  {Scalapino}},\ }\href {https://doi.org/10.1103/PhysRevB.84.180513} {\bibfield
   {journal} {\bibinfo  {journal} {Phys. Rev. B}\ }\textbf {\bibinfo {volume}
  {84}},\ \bibinfo {pages} {3} (\bibinfo {year} {2011})}\BibitemShut {NoStop}%
\bibitem [{\citenamefont {Maier}\ \emph
  {et~al.}(2005{\natexlab{b}})\citenamefont {Maier}, \citenamefont {Jarrell},
  \citenamefont {Schulthess}, \citenamefont {Kent},\ and\ \citenamefont
  {White}}]{maier_systematic_2005}%
  \BibitemOpen
  \bibfield  {author} {\bibinfo {author} {\bibfnamefont {T.~A.}\ \bibnamefont
  {Maier}}, \bibinfo {author} {\bibfnamefont {M.}~\bibnamefont {Jarrell}},
  \bibinfo {author} {\bibfnamefont {T.~C.}\ \bibnamefont {Schulthess}},
  \bibinfo {author} {\bibfnamefont {P.~R.~C.}\ \bibnamefont {Kent}},\ and\
  \bibinfo {author} {\bibfnamefont {J.~B.}\ \bibnamefont {White}},\ }\href
  {https://doi.org/10.1103/PhysRevLett.95.237001} {\bibfield  {journal}
  {\bibinfo  {journal} {Phys. Rev. Lett.}\ }\textbf {\bibinfo {volume} {95}},\
  \bibinfo {pages} {237001} (\bibinfo {year} {2005}{\natexlab{b}})}\BibitemShut
  {NoStop}%
\bibitem [{\citenamefont {Zhang}\ \emph {et~al.}(2023)\citenamefont {Zhang},
  \citenamefont {Lin}, \citenamefont {Moreo},\ and\ \citenamefont
  {Dagotto}}]{PhysRevB.108.L180510}%
  \BibitemOpen
  \bibfield  {author} {\bibinfo {author} {\bibfnamefont {Y.}~\bibnamefont
  {Zhang}}, \bibinfo {author} {\bibfnamefont {L.-F.}\ \bibnamefont {Lin}},
  \bibinfo {author} {\bibfnamefont {A.}~\bibnamefont {Moreo}},\ and\ \bibinfo
  {author} {\bibfnamefont {E.}~\bibnamefont {Dagotto}},\ }\href
  {https://doi.org/10.1103/PhysRevB.108.L180510} {\bibfield  {journal}
  {\bibinfo  {journal} {Phys. Rev. B}\ }\textbf {\bibinfo {volume} {108}},\
  \bibinfo {pages} {L180510} (\bibinfo {year} {2023})}\BibitemShut {NoStop}%
\bibitem [{\citenamefont {Troyer}\ and\ \citenamefont
  {Wiese}(2005)}]{troyer_computational_2005}%
  \BibitemOpen
  \bibfield  {author} {\bibinfo {author} {\bibfnamefont {M.}~\bibnamefont
  {Troyer}}\ and\ \bibinfo {author} {\bibfnamefont {U.}~\bibnamefont {Wiese}},\
  }\href {http://prola.aps.org/abstract/PRL/v94/i17/e170201} {\bibfield
  {journal} {\bibinfo  {journal} {Phys. Rev. Lett.}\ }\textbf {\bibinfo
  {volume} {94}},\ \bibinfo {pages} {170201} (\bibinfo {year}
  {2005})}\BibitemShut {NoStop}%
\bibitem [{\citenamefont {Liu}\ \emph {et~al.}(2016)\citenamefont {Liu},
  \citenamefont {Kaushal}, \citenamefont {Li}, \citenamefont {Bishop},
  \citenamefont {Wang}, \citenamefont {Johnston}, \citenamefont {Alvarez},
  \citenamefont {Moreo},\ and\ \citenamefont {Dagotto}}]{PhysRevE.93.063313}%
  \BibitemOpen
  \bibfield  {author} {\bibinfo {author} {\bibfnamefont {G.}~\bibnamefont
  {Liu}}, \bibinfo {author} {\bibfnamefont {N.}~\bibnamefont {Kaushal}},
  \bibinfo {author} {\bibfnamefont {S.}~\bibnamefont {Li}}, \bibinfo {author}
  {\bibfnamefont {C.~B.}\ \bibnamefont {Bishop}}, \bibinfo {author}
  {\bibfnamefont {Y.}~\bibnamefont {Wang}}, \bibinfo {author} {\bibfnamefont
  {S.}~\bibnamefont {Johnston}}, \bibinfo {author} {\bibfnamefont
  {G.}~\bibnamefont {Alvarez}}, \bibinfo {author} {\bibfnamefont
  {A.}~\bibnamefont {Moreo}},\ and\ \bibinfo {author} {\bibfnamefont
  {E.}~\bibnamefont {Dagotto}},\ }\href
  {https://doi.org/10.1103/PhysRevE.93.063313} {\bibfield  {journal} {\bibinfo
  {journal} {Phys. Rev. E}\ }\textbf {\bibinfo {volume} {93}},\ \bibinfo
  {pages} {063313} (\bibinfo {year} {2016})}\BibitemShut {NoStop}%
\bibitem [{\citenamefont {Hettler}\ \emph {et~al.}(1998)\citenamefont
  {Hettler}, \citenamefont {Tahvildar-Zadeh}, \citenamefont {Jarrell},
  \citenamefont {Pruschke},\ and\ \citenamefont
  {Krishnamurthy}}]{hettler_nonlocal_1998}%
  \BibitemOpen
  \bibfield  {author} {\bibinfo {author} {\bibfnamefont {M.~H.}\ \bibnamefont
  {Hettler}}, \bibinfo {author} {\bibfnamefont {A.~N.}\ \bibnamefont
  {Tahvildar-Zadeh}}, \bibinfo {author} {\bibfnamefont {M.}~\bibnamefont
  {Jarrell}}, \bibinfo {author} {\bibfnamefont {T.}~\bibnamefont {Pruschke}},\
  and\ \bibinfo {author} {\bibfnamefont {H.~R.}\ \bibnamefont
  {Krishnamurthy}},\ }\href {https://doi.org/10.1103/PhysRevB.58.R7475}
  {\bibfield  {journal} {\bibinfo  {journal} {Phys. Rev. B}\ }\textbf {\bibinfo
  {volume} {58}},\ \bibinfo {pages} {R7475} (\bibinfo {year}
  {1998})}\BibitemShut {NoStop}%
\bibitem [{\citenamefont {Gull}\ \emph
  {et~al.}(2011{\natexlab{a}})\citenamefont {Gull}, \citenamefont {Millis},
  \citenamefont {Lichtenstein}, \citenamefont {Troyer},\ and\ \citenamefont
  {Werner}}]{gull_continuous-time_2011}%
  \BibitemOpen
  \bibfield  {author} {\bibinfo {author} {\bibfnamefont {E.}~\bibnamefont
  {Gull}}, \bibinfo {author} {\bibfnamefont {A.~J.}\ \bibnamefont {Millis}},
  \bibinfo {author} {\bibfnamefont {A.~I.}\ \bibnamefont {Lichtenstein}},
  \bibinfo {author} {\bibfnamefont {M.}~\bibnamefont {Troyer}},\ and\ \bibinfo
  {author} {\bibfnamefont {P.}~\bibnamefont {Werner}},\ }\href
  {https://doi.org/10.1103/Revmodphys.83.349} {\bibfield  {journal} {\bibinfo
  {journal} {Rev. Mod. Phys.}\ }\textbf {\bibinfo {volume} {83}},\ \bibinfo
  {pages} {349} (\bibinfo {year} {2011}{\natexlab{a}})}\BibitemShut {NoStop}%
\bibitem [{\citenamefont {Gull}\ \emph
  {et~al.}(2011{\natexlab{b}})\citenamefont {Gull}, \citenamefont {Staar},
  \citenamefont {Fuchs}, \citenamefont {Nukala}, \citenamefont {Summers},
  \citenamefont {Pruschke}, \citenamefont {Schulthess},\ and\ \citenamefont
  {Maier}}]{gull_submatrix_2011}%
  \BibitemOpen
  \bibfield  {author} {\bibinfo {author} {\bibfnamefont {E.}~\bibnamefont
  {Gull}}, \bibinfo {author} {\bibfnamefont {P.}~\bibnamefont {Staar}},
  \bibinfo {author} {\bibfnamefont {S.}~\bibnamefont {Fuchs}}, \bibinfo
  {author} {\bibfnamefont {P.}~\bibnamefont {Nukala}}, \bibinfo {author}
  {\bibfnamefont {M.}~\bibnamefont {Summers}}, \bibinfo {author} {\bibfnamefont
  {T.}~\bibnamefont {Pruschke}}, \bibinfo {author} {\bibfnamefont
  {T.}~\bibnamefont {Schulthess}},\ and\ \bibinfo {author} {\bibfnamefont
  {T.}~\bibnamefont {Maier}},\ }\href
  {https://doi.org/10.1103/PhysRevB.83.075122} {\bibfield  {journal} {\bibinfo
  {journal} {Phys. Rev. B}\ }\textbf {\bibinfo {volume} {83}},\ \bibinfo
  {pages} {75122} (\bibinfo {year} {2011}{\natexlab{b}})}\BibitemShut {NoStop}%
\bibitem [{\citenamefont {Hähner}\ \emph {et~al.}(2020)\citenamefont
  {Hähner}, \citenamefont {Alvarez}, \citenamefont {Maier}, \citenamefont
  {Solcà}, \citenamefont {Staar}, \citenamefont {Summers},\ and\ \citenamefont
  {Schulthess}}]{hahner_dca_2020}%
  \BibitemOpen
  \bibfield  {author} {\bibinfo {author} {\bibfnamefont {U.~R.}\ \bibnamefont
  {Hähner}}, \bibinfo {author} {\bibfnamefont {G.}~\bibnamefont {Alvarez}},
  \bibinfo {author} {\bibfnamefont {T.~A.}\ \bibnamefont {Maier}}, \bibinfo
  {author} {\bibfnamefont {R.}~\bibnamefont {Solcà}}, \bibinfo {author}
  {\bibfnamefont {P.}~\bibnamefont {Staar}}, \bibinfo {author} {\bibfnamefont
  {M.~S.}\ \bibnamefont {Summers}},\ and\ \bibinfo {author} {\bibfnamefont
  {T.~C.}\ \bibnamefont {Schulthess}},\ }\href
  {https://doi.org/10.1016/j.cpc.2019.01.006} {\bibfield  {journal} {\bibinfo
  {journal} {Comp. Phys. Comm.}\ }\textbf {\bibinfo {volume} {246}},\ \bibinfo
  {pages} {106709} (\bibinfo {year} {2020})}\BibitemShut {NoStop}%
\end{thebibliography}%


\begin{thebibliography}{2}%
\makeatletter
\providecommand \@ifxundefined [1]{%
 \@ifx{#1\undefined}
}%
\providecommand \@ifnum [1]{%
 \ifnum #1\expandafter \@firstoftwo
 \else \expandafter \@secondoftwo
 \fi
}%
\providecommand \@ifx [1]{%
 \ifx #1\expandafter \@firstoftwo
 \else \expandafter \@secondoftwo
 \fi
}%
\providecommand \natexlab [1]{#1}%
\providecommand \enquote  [1]{``#1''}%
\providecommand \bibnamefont  [1]{#1}%
\providecommand \bibfnamefont [1]{#1}%
\providecommand \citenamefont [1]{#1}%
\providecommand \href@noop [0]{\@secondoftwo}%
\providecommand \href [0]{\begingroup \@sanitize@url \@href}%
\providecommand \@href[1]{\@@startlink{#1}\@@href}%
\providecommand \@@href[1]{\endgroup#1\@@endlink}%
\providecommand \@sanitize@url [0]{\catcode `\\12\catcode `\$12\catcode
  `\&12\catcode `\#12\catcode `\^12\catcode `\_12\catcode `\%12\relax}%
\providecommand \@@startlink[1]{}%
\providecommand \@@endlink[0]{}%
\providecommand \url  [0]{\begingroup\@sanitize@url \@url }%
\providecommand \@url [1]{\endgroup\@href {#1}{\urlprefix }}%
\providecommand \urlprefix  [0]{URL }%
\providecommand \Eprint [0]{\href }%
\providecommand \doibase [0]{https://doi.org/}%
\providecommand \selectlanguage [0]{\@gobble}%
\providecommand \bibinfo  [0]{\@secondoftwo}%
\providecommand \bibfield  [0]{\@secondoftwo}%
\providecommand \translation [1]{[#1]}%
\providecommand \BibitemOpen [0]{}%
\providecommand \bibitemStop [0]{}%
\providecommand \bibitemNoStop [0]{.\EOS\space}%
\providecommand \EOS [0]{\spacefactor3000\relax}%
\providecommand \BibitemShut  [1]{\csname bibitem#1\endcsname}%
\let\auto@bib@innerbib\@empty
\bibitem [{\citenamefont {Jarrell}\ \emph {et~al.}(2001)\citenamefont
  {Jarrell}, \citenamefont {Maier}, \citenamefont {Huscroft},\ and\
  \citenamefont {Moukouri}}]{jarrell_quantum_2001}%
  \BibitemOpen
  \bibfield  {author} {\bibinfo {author} {\bibfnamefont {M.}~\bibnamefont
  {Jarrell}}, \bibinfo {author} {\bibfnamefont {T.}~\bibnamefont {Maier}},
  \bibinfo {author} {\bibfnamefont {C.}~\bibnamefont {Huscroft}},\ and\
  \bibinfo {author} {\bibfnamefont {S.}~\bibnamefont {Moukouri}},\ }\href
  {https://doi.org/10.1103/PhysRevB.64.195130} {\bibfield  {journal} {\bibinfo
  {journal} {Physical Review B}\ }\textbf {\bibinfo {volume} {64}},\ \bibinfo
  {pages} {195130} (\bibinfo {year} {2001})},\ \bibinfo {note} {publisher:
  American Physical Society}\BibitemShut {NoStop}%
\bibitem [{\citenamefont {Maier}\ \emph {et~al.}(2005)\citenamefont {Maier},
  \citenamefont {Jarrell}, \citenamefont {Pruschke},\ and\ \citenamefont
  {Hettler}}]{maier_quantum_2005}%
  \BibitemOpen
  \bibfield  {author} {\bibinfo {author} {\bibfnamefont {T.}~\bibnamefont
  {Maier}}, \bibinfo {author} {\bibfnamefont {M.}~\bibnamefont {Jarrell}},
  \bibinfo {author} {\bibfnamefont {T.}~\bibnamefont {Pruschke}},\ and\
  \bibinfo {author} {\bibfnamefont {M.}~\bibnamefont {Hettler}},\ }\href
  {http://publish.aps.org/} {\bibfield  {journal} {\bibinfo  {journal} {Reviews
  of Modern Physics}\ }\textbf {\bibinfo {volume} {77}},\ \bibinfo {pages}
  {1027} (\bibinfo {year} {2005})}\BibitemShut {NoStop}%
\end{thebibliography}%

\end{document}


\title{Supplemental Information: Interlayer Pairing in Bilayer Nickelates}

\date{\today}

\author{Thomas A. Maier} \affiliation{Computational Sciences and Engineering Division, Oak Ridge National Laboratory, Oak Ridge, Tennessee 37831, USA}
\author{Peter Doak} \affiliation{Computational Sciences and Engineering Division, Oak Ridge National Laboratory, Oak Ridge, Tennessee 37831, USA}
\author{Ling-Fang Lin} \affiliation{Department of Physics and Astronomy, University of Tennessee, Knoxville, Tennessee 37996, USA}
\author{Yang Zhang} \affiliation{Department of Physics and Astronomy, University of Tennessee, Knoxville, Tennessee 37996, USA}
\author{Adriana Moreo} \affiliation{Department of Physics and Astronomy, University of Tennessee, Knoxville,  Tennessee 37996, USA} \affiliation{Materials Science and Technology Division, Oak Ridge National Laboratory, Oak Ridge, Tennessee 37831, USA}
\author{Elbio Dagotto} \affiliation{Department of Physics and Astronomy, University of Tennessee, Knoxville,  Tennessee 37996, USA} \affiliation{Materials Science and Technology Division, Oak Ridge National Laboratory, Oak Ridge, Tennessee 37831, USA}

\maketitle

In order to calculate the pair-field susceptibility (Eq.~(1) in the main text), we follow the usual DCA formalism described in Refs.~\cite{jarrell_quantum_2001, maier_quantum_2005} to calculate susceptibilities for the lattice in the thermodynamic limit. This requires a calculation of the four-point two-particle Green's function
\begin{equation}
    G^{pp}_{\ell_1\ell_2\ell_3\ell_4}(x_1,x_2;x_3,x_4) = -\langle {\cal T}_\tau c^{\phantom\dagger}_{\ell_1\uparrow}(x_1)c^{\phantom\dagger}_{\ell_2\downarrow}(x_2)c^{\dagger}_{\ell_4\downarrow}(x_4)c^{\dagger}_{\ell_3\uparrow}(x_3)\rangle
\end{equation}
After Fourier-transforming on both the space and time variables, restricting to zero momentum and frequency transfer $Q=0$, and the usual DCA coarse-graining of momentum space (for details see \cite{maier_quantum_2005}), one obtains the coarse-grained two-particle Green's function ${\bar G}^{pp}_{\ell_1\ell_2\ell_3\ell_4}(K,K')$ with $K=(\bm K, i\omega_n)$ and $K'=(\bm K', i\omega_{n'})$ where $\bm K$ are the cluster momenta and $\omega_n$ the fermionic Matsubara frequencies.
The pair-field susceptibility in Eq.~(1) in the main text is then obtained as
\begin{equation}
  P_\alpha(T) = \frac{T^2}{N_c^2}\sum_{K,K'}\sum_{\ell_1\ell_2\ell_3\ell_4}g_\alpha^{\ell_1\ell_2}(\bm K) {\bar G}^{pp}_{\ell_1\ell_2\ell_3\ell_4}(K,K')g_\alpha^{\ell_3\ell_4}(\bm K')\,.
\end{equation}
${\bar G}^{pp}_{\ell_1\ell_2\ell_3\ell_4}$ may be written as a Bethe-Salpeter equation
\begin{eqnarray}
    {\bar G}^{pp}_{\ell_1\ell_2\ell_3\ell_4}(K,K') = \bar{G}^{pp,0}_{\ell_1\ell_2\ell_3\ell_4}(K)\delta_{K,K'} + G^{pp,\Gamma}_{\ell_1\ell_2\ell_3\ell_4}(K,K')
\end{eqnarray}
with the coarse-grainined zeroth order (in the interaction) contribution
\begin{equation}
    {\bar G}^{pp,0}_{\ell_1\ell_2\ell_3\ell_4}(K) = \frac{N_c}{N}\sum_{\bm k'}G_{\ell_1\ell_3}(\bm K+\bm k', i\omega_n)G_{\ell_2\ell_4}(-\bm K-\bm k', -i\omega_{n'})
\end{equation}
and the higher-order contributions from the interaction term (Eq.~(8) in the main text)
\begin{eqnarray}
    {\bar G}^{pp,\Gamma}_{\ell_1\ell_2\ell_3\ell_4}(K,K') = \frac{T}{N_c}\sum_{\ell_5\ell_6\ell_7\ell_8} \bar{G}^{pp,0}_{\ell_1\ell_2\ell_5\ell_6}(K) \Gamma_{\ell_5\ell_6\ell_7\ell_8}(K, K') \bar{G}^{pp,0}_{\ell_7\ell_8\ell_3\ell_4}(K')\,.
\end{eqnarray}
Here, $G_{\ell\ell'}(\bm k, i\omega_n)$ is the single-particle Green's function and $\Gamma_{\ell_1\ell_2\ell_3\ell_4}(K, K')$ is the reducible four-point vertex. ${\bar G}^{pp,\Gamma}_{\ell_1\ell_2\ell_3\ell_4}$ is illustrated diagrammatically in Fig.~3a of the main text.
The pair structures $\phi^\alpha_{\ell\ell'}(\bm r)$ shown in Fig.~4 of the main text are obtained after Fourier-transforming to real space the eigenvectors $\phi^\alpha_{\ell\ell'}(\bm K)$ of the eigenvalue equation
\begin{equation}
    \sum_{\bm K',\ell_3\ell_4}\left[T\sum_{\omega_n,\omega_{n'}} {\bar G}^{pp,\Gamma}_{\ell_1\ell_2\ell_3\ell_4}({\bm K},\omega_n,{\bm K'},\omega_{n'})\right]\phi^\alpha_{\ell_3\ell_4}({\bm K'}) = \lambda^{\alpha} \phi^\alpha_{\ell_1\ell_2}({\bm K})\,.
\end{equation}
Here we have summed ${\bar G}^{pp,\Gamma}_{\ell_1\ell_2\ell_3\ell_4}$ over frequency since we are only interested in the even frequency singlet states.


\bibliography{supplement}